\documentclass[12pt,journal,onecolumn,draftcls]{IEEEtran}

\usepackage{cite, cancel}      

\usepackage{graphicx}  

\usepackage{psfrag}    

\usepackage{subfigure} 

\usepackage{url}       

\usepackage{stfloats}  

\usepackage{amsmath}   
\usepackage{array}
 \makeatletter
 \let\NAT@parse\undefined
 \makeatother
\usepackage{lipsum}

\usepackage{times}
\usepackage{epsfig,graphicx,color,pstricks}
\usepackage{amsmath,amssymb,amsthm,amsbsy,amsfonts,latexsym}
\usepackage{bm}
\usepackage{psfrag}
\usepackage[mathscr]{eucal}
\usepackage{cite}
\usepackage[all]{xy}
\usepackage{subfigure}
\usepackage{graphics}

\usepackage{multicol}
\usepackage{lipsum}

\newtheorem{thm}{Theorem}
\newtheorem{cor}[thm]{Corollary}

\newtheorem{prop}[thm]{Proposition}

\theoremstyle{rem}
\newtheorem{rem}{Remark}

\newcommand{\eu}{\mathrm{e}}

\newcommand{\mud}{{\mathsf{I_d}}}
\newcommand{\mug}{{\mathsf{I_g}}}

\newcommand{\points}{{\mathsf{N_d}}}
\newcommand{\supp}{{\mathsf{supp}}}
\newcommand{\snr}{{\mathsf{SNR}}}
\newcommand{\inr}{{\mathsf{INR}}}

\newcommand{\co}{{\mathsf{co}}}

\newcommand{\gaptoclassicalIC}{{{$\frac{1}{2}\log\left(12\pi\eu\right)\approx3.34$}}} 

\begin{document}

\title{On the Two-user Interference Channel with Lack of Knowledge of the Interference Codebook at one Receiver}
 

\author{Alex Dytso, Daniela Tuninetti, and Natasha Devroye%
\thanks{%
The authors are with the Electrical and Computer Engineering Department of the University of Illinois at Chicago, Chicago, IL 60607 USA (e-mail: odytso2, danielat, devroye @uic.edu). 

The results in this paper were presented in part at the  IEEE International Symposium on Information Theory, Istanbul, Turkey, July 2013 and at the IEEE Information Theory and Applications Workshop, San Diego, USA, February 2014. 

The work of the authors was partially funded by NSF under award number 1017436. The contents of this article are solely the responsibility of the authors and do not necessarily represent the official views of the NSF.

Copyright (c) 2013 IEEE. Personal use of this material is permitted.  However, permission to use this material for any other purposes must be obtained from the IEEE by sending a request to pubs-permissions@ieee.org.%
}
}
\maketitle

%

\maketitle

\begin{abstract}
In multi-user information theory it is often assumed that every node in the network possesses all codebooks used in the network. This assumption may be impractical in distributed ad-hoc, cognitive or heterogeneous networks. 
This work considers the two-user Interference Channel with one {\em Oblivious Receiver} (IC-OR), i.e., one receiver lacks knowledge of the interfering cookbook while the other receiver knows both codebooks.
The paper asks whether, and if so how much, the channel capacity of the IC-OR is reduced compared to that of the classical IC where both receivers know all codebooks. 
A novel outer bound is derived and shown to be achievable to within a  gap 
for the class of injective semi-deterministic IC-ORs; the gap is shown to be zero for injective fully deterministic IC-ORs.
%
An exact capacity result is shown for the general memoryless IC-OR when the non-oblivious receiver experiences very strong interference. 
%
%
For the linear deterministic IC-OR that models the Gaussian noise channel at high SNR, 
non i.i.d. Bernoulli(1/2) input bits are shown to achieve points not achievable by i.i.d. Bernoulli(1/2) input bits used in the same achievability scheme.  
For the real-valued Gaussian IC-OR the gap is shown to be at most $1/2$~bit per channel use, even though the set of optimal input distributions for the derived outer bound could not be determined. { Towards understanding the Gaussian IC-OR, an achievability strategy is evaluated in which the input alphabets at the non-oblivious transmitter are a mixture of discrete and Gaussian random variables,  where the cardinality of the discrete part is appropriately chosen as a function of the channel parameters. Surprisingly, as the oblivious receiver intuitively should not be able to ``jointly decode'' the intended and interfering messages (whose codebook is unavailable), it is shown  that with this choice of input, the capacity region of the symmetric Gaussian IC-OR
is to within \gaptoclassicalIC~bits (per channel use per user) of an outer bound for the classical Gaussian IC with full codebook  knowledge at both receivers. }
%
\end{abstract}

\section{Introduction}
A classical assumption in multi-user information theory is that each node in the network possesses knowledge of the codebooks used by every other node.  However, such an assumption might not be practical in heterogeneous, cognitive, distributed or dynamic networks.
For example, in very large  ad-hoc networks, where nodes enter and leave at will, it might not be practical for new nodes to learn the codebooks of old nodes and vice-versa. In cognitive radio scenarios, where new cognitive systems coexist with legacy systems, requiring the legacy systems to know the codebooks of the new cognitive systems might not be viable. 
This motivates the study of networks where each node possesses only a subset of the codebooks used in the network. We will refer to such systems as networks with \emph{partial codebook knowledge} and to nodes with only knowledge of a subset of the codebooks as \emph{oblivious}  nodes. 

To the best of our knowledge, systems with oblivious terminals were first introduced in \cite{sand_decentr_proces}. In \cite{sand_decentr_proces} lack of codebook knowledge was modeled by using {\it codebook indices}, which index the random encoding function that maps the messages to the codewords. If a node has codebook knowledge it knows the index (or instance) of the random encoding function used; else it does not and the codewords essentially look like the symbols were produced in an independent, identically distributed (i.i.d.) fashion from a given distribution. 
In \cite{simion_codebook} and \cite{CodebookYEner} this concept of partial codebook knowledge was extended to model oblivious relays and capacity results were derived.
However, as pointed out in \cite[Section III.A]{simion_codebook} and \cite[Remark 5]{CodebookYEner}, these capacity regions are ``non-computable'' in the sense that it is not known how to find the optimal input distribution in general. In particular, the capacity achieving input distribution for the practically relevant Gaussian noise channel remains an open problem.  

We make progress on this front by demonstrating that certain rates are achievable for the Gaussian noise interference channel with oblivious receivers (G-IC-OR) 
through the evaluation of a simplified Han-Kobayashi scheme \cite{H+K} in which joint decoding of the intended and interfering messages is not required at the oblivious receiver.
{
The major contribution of this work is the realization that Gaussian inputs perform poorly in the proposed achievable region. We therefore propose to use a class of inputs that we termed
{\it mixed inputs}. A mixed input is random variable that is a mixture of a continuous and a discrete part, such as for example a Gaussian random variable and a uniformly distributed random variable on an equally spaced set of discrete points.
We then properly design the distribution of the mixed input as a function of the channel parameters. 
}

We are not the first to consider discrete inputs for Gaussian noise channels. 
In \cite{PAMozarow} the authors considered the point-to-point power-constrained Gaussian noise channel and derived lower bounds on the achievable rate when the input is contained to be an equally spaced Pulse Amplitude Modulation (PAM) in which each each point is used with equal probability; such an input was shown to be optimal to within 0.41~bits per channel use \cite[eq.(9)]{PAMozarow}. 
As pointed out in \cite{ShamaiShannonLecture}, already in 1948 Claude Shannon in the unpublished work \cite{ShanonConstallationsUnpublished} argued the asymptotically optimality of a PAM input for the point-to-point power-constrained Gaussian noise channel.

In \cite[Theorems 6 and 7]{WuVerduDOFAllerton10}, the authors {\it asymptotically} characterized the optimal input distribution over $N$ masses at high and low $\snr$, respectively, for a point-to-point power-constrained Gaussian noise channel by assuming that $N$ is not dependent on $\snr$. 

For the purpose of this work, these bounds cannot be used, as 
1) these bounds are optimized for a specific $\snr$ while we shall need to lower bound the rate achievable by a discrete input at multiple receivers each characterized by a different $\snr$; 2) we need a {\it firm} bound that holds at all finite $\snr$; and 3) we need to properly choose $N$ as a function of $\snr$, a question posed but left open in \cite{WuVerduDOFAllerton10}.

The sub-optimality of Gaussian inputs for Gaussian noise channels has been observed before. 
Past work on the asynchronous IC \cite{totAsynchIC, CoordinateSystemAbbe} showed that non-Gaussian inputs may outperform i.i.d. Gaussian inputs by using local perturbations of an i.i.d. Gaussian input: \cite[Lemma 3]{totAsynchIC} considers a fourth order approximation of mutual information, while \cite[Theorem 4]{CoordinateSystemAbbe} uses perturbations in the direction of Hermite polynomials of order larger than three. In both cases the input distribution is assumed to have a density, though \cite[Fig. 1]{totAsynchIC} numerically shows the performance of a ternary PAM input as well. For the cases considered in \cite{totAsynchIC, CoordinateSystemAbbe}, the improvement over i.i.d. Gaussian inputs shows in the decimal digits of the achievable rates; it is hence not clear that perturbed continuous Gaussian inputs as in \cite{totAsynchIC, CoordinateSystemAbbe} can actually provide DoF gains over Gaussian inputs (note that a strict DoF gain implies an unbounded rate gain as $\snr$ increases) which we seek in this work.
In a way this work follows the philosophy of  \cite{inPraiseOfBadCodes}: the main idea is to use sub-optimal point-to-point codes in which the reduction in achievable rates is more than compensated by the decrease in the interference created at the other users.

The rest of the paper is organized as follows.
The general memoryless IC-OR channel model is introduced in Section \ref{sec:channelModel}, together with the special class of {\em injective semi-deterministic} IC-ORs (ISD-IC-OR) of which the Gaussian noise channel is an example.  Our main results are:
\begin{enumerate}

\item
In Section \ref{sec:ch:main}, in Theorem~\ref{thm:outerbound} we derive a novel outer bound that incorporates this partial codebook knowledge explicitly. In this bound, the single rate bounds are valid for a general memoryless IC-OR while the sum-rate bound is valid for the ISD-IC-OR only.

\item
In Section \ref{sec:ch:capacity} we demonstrate a series of capacity and approximate capacity results for various regimes and classes of IC-OR. Specifically, by using the achievable region in Proposition~\ref{prop:innerbound} we prove:
(a) in  Theorem~\ref{thm:very strong cap memoryless} we obtain the capacity region for the general memoryless IC-OR in very strong interference at the non-oblivious receiver, 
(b) in Theorem~\ref{thm:gap} we demonstrate the capacity region to within a gap for the ISD-IC-OR, and
(c) in Corollary~\ref{cor:cap:Deterministic} we show that for the injective fully deterministic IC-OR the gap is zero.

\item
In Section~\ref{sec:get no gap}, we look at the practically relevant G-IC-OR and its corresponding Linear Deterministic Approximation (LDA-IC-OR) in the spirit of \cite{AvestDeterministic}, which models the G-IC-OR at high SNR. 
Surprisingly, for the LDA-IC-OR we numerically demonstrate that for the proposed achievable scheme in Proposition~\ref{prop:innerbound},  i.i.d. Bernoulli(1/2) input bits (known to be optimal for the LDA-IC with full codebook knowledge \cite{bresler_tse}) are outperformed by other (correlated and non-uniform) input distributions.

\item  In Section~\ref{sec:ch:Gaussian}, for the G-IC-OR, we show  in Corollary \ref{gap:const:gaus} that our inner and outer bounds that are to within $1/2$~bit (per channel use per user) of one another. However, similarly to prior work on oblivious models, we are not able to find the set of input distributions that exhaust the outer bound in Theorem~\ref{thm:outerbound}, in particular we cannot argue whether i.i.d. Gaussian inputs exhaust the outer bound.

Inspired by the results for the LDA-IC-OR,

we numerically show that a larger sum-capacity is attainable by using a discrete input at the non-oblivious transmitter than by selecting i.i.d. Gaussian inputs, or using time-division, or treating interference as noise, in the strong interference regime at high SNR. 
This suggests that the penalty for the lack of codebook knowledge is not as severe as one might initially expect.

\item
For the remainder of the paper we consider the G-IC-OR, and demonstrate that even with partial codebook knowledge we are able to achieve to within \gaptoclassicalIC~bits per channel use of the {\it symmetric capacity region} of the G-IC with full codebook knowledge through the use of mixed inputs.\footnote{The restriction to the symmetric case, i.e., same direct links and same interference links, is just to reduce the number of parameters in our derivations. We strongly believe that an approximate capacity result (to within a constant gap) can be shown for the general asymmetric case, albeit through more tedious computations than those reported here.}
The main tool, to derive the symmetric capacity to within a constant gap is Theorem~\ref{thm:lowbound},
which is the lower bound from \cite{PAMozarow} on the mutual information achievable by a PAM input on a point-to-point Gaussian noise channel.
With this tool, in Section \ref{sec:achScheme} in Theorems~\ref{thm:sec:achScheme G-IC-OR} and~\ref{thm:sec:achScheme G-IC-OR:Ach2}, we evaluate the achievable rate region presented in Proposition~\ref{prop:innerbound} for the G-IC-OR when the non-oblivious transmitter uses either a PAM input or a mixed input that comprises a Gaussian component and a PAM component. Corollaries~\ref{thm:ach:gDoF:Region:strongInteference} and~\ref{thm:ach:gDoF:Region:weakInteference} provide the gDoF characterization of the achievable regions in Theorems~\ref{thm:sec:achScheme G-IC-OR} and~\ref{thm:sec:achScheme G-IC-OR:Ach2}.

\item
In past work on networks with oblivious nodes no performance guarantees were provided as the capacity regions could not be evaluated.
In Section \ref{sec:gDoFach} we study the generalized degrees of freedom (gDoF) achievable with mixed inputs. In Theorem~\ref{thm:GDOFregion}, we show that mixed inputs achieve the gDoF of the classical G-IC, hence implying that there is no loss in performance due to lack of codebooks in a gDoF sense /  at high SNR. 
This is quite surprising considering that the oblivious receiver cannot perform joint decoding of the two messages, which is optimal for the classical G-IC in the strong and very strong interference regimes. 

\item
Finally, in Section \ref{sec:finiteSNR} we turn our attention to the finite $\snr$ regime and  in Theorem \ref{thm:cosntant gap} we show that the capacity of the symmetric G-IC-OR is within \gaptoclassicalIC~bits per channel use of the outer bound to the capacity region of the classical symmetric G-IC.

\end{enumerate}

We conclude the paper with some final remarks and future directions in Section \ref{sec:Conclusion}.
Some proofs are reported in the Appendix.

\section{Channel Model}
\label{sec:channelModel}

\subsection{Notation} 
\label{sec:notation}
We adopt the following notation convention: 
\begin{itemize}
\item
Lower case variables are instances of upper case random variables which take on values in calligraphic alphabets.

\item
Throughout the paper $\log(\cdot)$ denotes logarithms in base 2.
 
\item
$[n_1:n_2]$ is the set of integers from $n_1$ to $n_2 \geq n_1$.

\item
$Y^{j}$ is a vector of length $j$ with components $(Y_1,\ldots,Y_j)$.

\item
We let $\delta(\cdot)$ denote the the Dirac delta function.  

\item
If $A$ is a random variable (r.v.) we denote its support by $\supp(A)$.  \item
The symbol $|\cdot|$ may denote different things: 
$|\mathcal{A}|$ is the cardinality of the set $\mathcal{A}$, 
$|X|$ is the cardinality of $\supp(X)$ of the r.v. $X$ , or
$|x|$ is the absolute value of the real-valued $x$.

\item
For $x\in\mathbb{R}$ we let
$\left \lfloor x \right\rfloor$ denote the largest integer not greater than $x$.

\item
For $x\in\mathbb{R}$ we let $[x]^{+} :=\max(x,0)$ and $\log^{+}(x) :=[\log(x)]^+$.

\item
The functions $\mug(x)$, $\mud(n,x)$ and $\points(x)$, for $n\in\mathbb{N}$ and $x\in\mathbb{R}^+$, are defined as
\begin{align}
\mug(x) &:=\frac{1}{2}\log(1+x),
\\
\mud(n,x) &:=
\left[\frac{1}{2}\log(1+\min(n^2-1,x))-\frac{1}{2}\log\left(\frac{\pi\eu}{3}\right)\right]^+,
\\
\points(x)&:=\left \lfloor \sqrt{1+x} \right\rfloor,
\end{align}
where the subscript $\mathsf{d}$ reminds the reader that discrete inputs are involved, while $\mathsf{g}$ that Gaussian inputs are involved.

\item
$\mathcal{N}(\mu,\sigma^2)$ denotes a real-valued Gaussian r.v. with mean $\mu$ and variance $\sigma^2$.

\item
$\mathsf{Unif}([n_1:n_2])$ denotes the uniform distribution over the set $[n_1:n_2]$.

\item
$\mathsf{Bernoulli}(p)$ denotes the Bernoulli distribution with parameter $p\in[0,1]$.

\item
$X \sim \mathsf{PAM}(N)$ denotes the uniform distribution over a zero-mean Pulse Amplitude Modulation (PAM) constellation with $|X|=N$ points and unit-energy. 
\item
$\co(\cdot)$ denotes  the convex closure operator.

\end{itemize}

\subsection{General Memoryless IC-OR}
\label{sec:channelModel:general}

\begin{figure}
 \center
 \includegraphics[width=12cm]{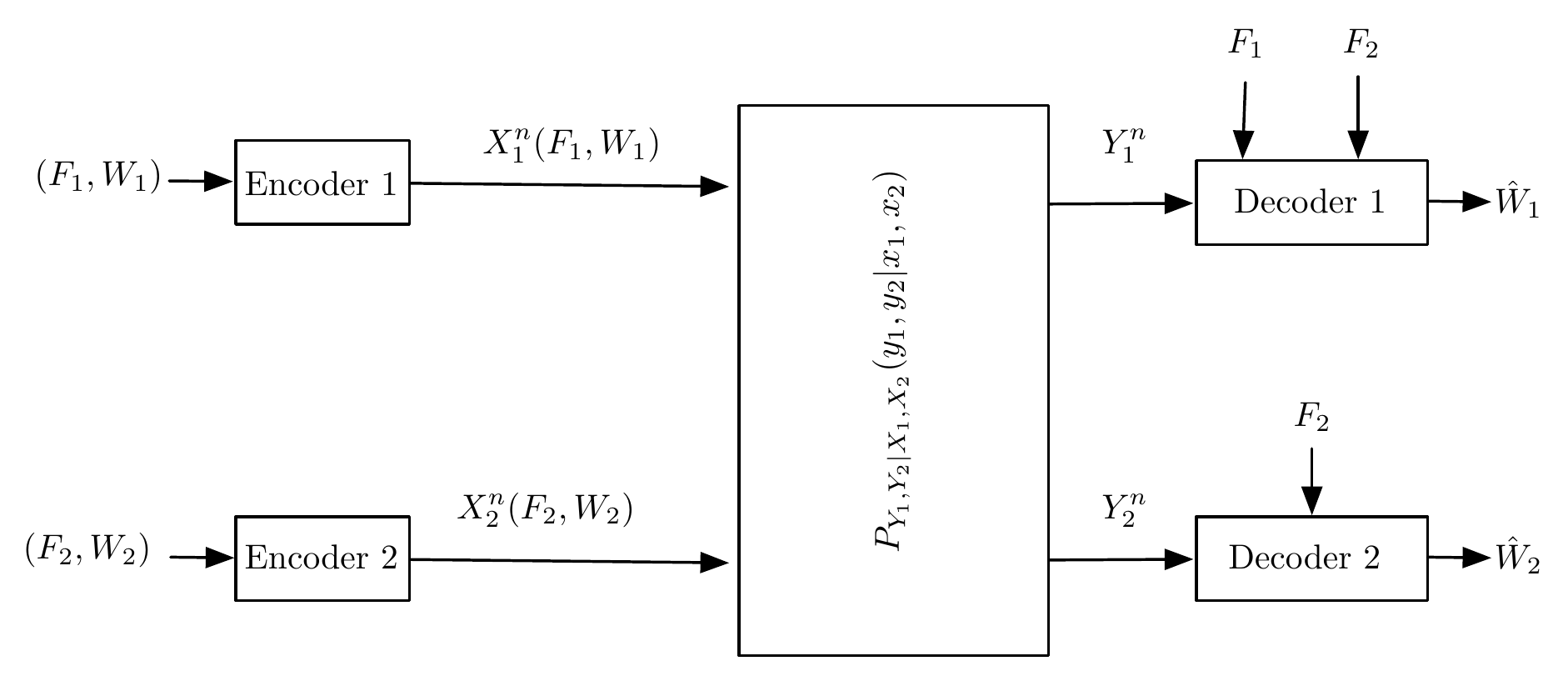}
 \caption{The IC-OR, where $F_1$ and $F_2$ represent codebook indices known to one or both receivers.}  \label{fig:channel:model}
\end{figure}

An IC-OR consists of the two-user memoryless interference channel $(\mathcal{X}_1,\mathcal{X}_2,P_{Y_1 Y_2|X_1 X_2},\mathcal{Y}_1,\mathcal{Y}_2)$ where receiver~2 is oblivious of transmitter~1's codebook. We use the terminology ``codebook'' to denote a set of codewords and the (one-to-one) mapping of the messages to these codewords.  We model lack of codebook knowledge as in \cite{sand_decentr_proces}, where transmitters use randomized encoding functions, which are indexed by a message index and a ``codebook index'' ($F_1$ and $F_2$ in Fig. \ref{fig:channel:model}). An oblivious receiver is unaware of the ``codebook index'' ($F_1$ is not given to decoder~2 in Fig. \ref{fig:channel:model}) and hence does not know how codewords are mapped to messages. The basic modeling assumption  is that without the knowledge of the codebook index a codeword looks unstructured. More formally, by extending \cite[Definition 2]{simion_codebook}, a ($2^{nR_1},2^{nR_2},n$) code for the IC-OR with enabled time sharing is a six-tuple $(P_{F_1|Q^n},\sigma_1^n,\phi_1^n,P_{F_2|Q^n},\sigma_2^n,\phi_2^n)$, where the distribution $P_{F_i|Q^n}$, $i\in [1:2]$, is over a finite alphabet  $\mathcal{F}_i$ conditioned on the time-sharing sequences $q^n$ from some finite alphabet $\mathcal{Q}$, and where the encoders $\sigma_i^n$ and the decoders $\phi_i^n$,  are mappings 
\begin{align*}
&\sigma_1^n: [1:2^{nR_1}] \times [1:|\mathcal{F}_1|] \rightarrow \mathcal{X}_1^n,\\
&\sigma_2^n: [1:2^{nR_2}] \times [1:|\mathcal{F}_2|] \rightarrow \mathcal{X}_2^n,\\
&\phi_1^n: [1:|\mathcal{F}_1|]  \times [1:|\mathcal{F}_2|] \times \mathcal{Y}_1^n \rightarrow [1:2^{nR_1}],\\
&\phi_2^n: [1:|\mathcal{F}_2|] \times \mathcal{Y}_2^n \rightarrow [1:2^{nR_2}].
\end{align*}
Moreover, when user 1's codebook index is unknown at decoder 2,  the encoder $\sigma_1^n$ and the distribution $P_{F_1|Q^n}$ must satisfy 
\begin{align}
  &\mathbb{P}[X_1^n = x_1^n | Q^n = q^n] \nonumber
\\&=\sum_{w_1=1}^{2^{nR_1}} \sum_{f_1=1}^{|\mathcal{F}_1|}P_{F_1|Q^n}(f_1|q^n) \ 2^{-nR_1} \ \delta\big(x_1^n - \sigma_1^n(w_1,f_1)\big) \nonumber
\\&=\prod_{t\in[1:n]} P_{X_1|Q}(x_{1t}|q_t)
\label{eq:osvaldo:eq},
\end{align}
according to some distribution $P_{X_1|Q}$. 
In other words, when averaged over the probability of selecting a given codebook and over a uniform distribution on the message set, the transmitted codeword conditioned on any time sharing sequence has an i.i.d. distribution according to some distribution $P_{X_1|Q}$. We refer the reader to \cite[Remark 1]{simion_codebook} for further justifications of the condition in~\eqref{eq:osvaldo:eq}. 

A non-negative rate pair $(R_1,R_2)$ is said to be achievable if there exist a sequence of encoding functions
$
\sigma_1^n(W_1,F_1), \
\sigma_2^n(W_2,F_2), \
$
and decoding functions
$
\phi_1^n(Y_1^n,F_1,F_2), \
\phi_2^n(Y_2^n,F_2), 
$

such that  the average probability of error satisfies
$\max_{i \in [1:2]}\mathbb{P}[\widehat{W}_i \neq W_i] \rightarrow 0$ as $n \rightarrow +\infty.$
The capacity region is defined as the convex closure of all achievable rate pairs $(R_1,R_2)$ \cite{elgamalkimbook}.

.

\subsection{Injective Semi-Deterministic IC-OR}
\label{sec:channelModel:isd}

For a general memoryless IC-OR, no restrictions are imposed on the transition probability $P_{Y_1 Y_2|X_1 X_2}$.
The ISD-IC-OR is a special IC-OR with transition probability
\begin{align}
  P_{Y_1 Y_2|X_1 X_2}(y_1,y_2|x_1,x_2)
  &= \sum_{t_1,t_2}P_{T_1|X_1}(t_1|x_1)P_{T_2|X_2}(t_2|x_2)
\nonumber\\&\cdot
  \delta\big(y_1 - g_{1}(x_1,t_2)\big)
\ \delta\big(y_2 - g_{2}(x_2,t_1)\big),
\label{eq:ISD def}
\end{align}
for some memoryless transition probabilities $P_{T_1|X_1}$ and $P_{T_2|X_2}$, and some deterministic functions $g_{1}(\cdot,\cdot)$ and $g_{2}(\cdot,\cdot)$ that are injective when their first argument is held fixed~\cite{Telatar_Tse_inject}.
The ISD property implies that 
\begin{align}
H(Y_1|X_1)=H(T_2) \ \text{and} \ H(Y_2|X_2)=H(T_1), \quad \forall P_{X_1 X_2}=P_{X_1}P_{X_2},
\label{eq:ISD implications}
\end{align}
or in other words that the $T_u$ is a deterministic function of the pair $(Y_u,X_u)$, $u\in[1:2]$.
For channels with continuous alphabets, the summation in~\eqref{eq:ISD def} should be replaced with an integral and the discrete entropies in~\eqref{eq:ISD implications} with the differential entropies.

\section{Outer Bounds} 
\label{sec:ch:main}
In this section we present novel outer bounds for the IC-OR. In particular, we derive the single rate bounds that are valid for a general memoryless IC-OR and a sum-rate bound that is valid for the ISD-IC-OR only.

We begin by proving a property of the output distributions that is key to 
deriving single-letter expressions in our outer bounds;
this property holds for a general memoryless IC-OR.

\begin{prop}
\label{prop:product}
The output of the oblivious decoder has a product distribution conditioned on the signal whose codebook is known, that is,
\begin{align*}
P_{Y_2^n|X_2^n,F_2}(y_2^n|x_2^n,f_2)=\prod^n_{i=1} P_{Y_{2i}|X_{2i}}(y_{2i}|x_{2i}).
\end{align*}
which implies
\begin{align*}
H(Y_2^n|X_2^n,F_2)
  &=\sum_{i=1}^nH(Y_{2i}|X_{2i})
\\&\stackrel{\text{\rm for ISD-IC-OR}}{=}\sum_{i=1}^nH(T_{1i}).
\end{align*}
\end{prop}
\begin{IEEEproof}[Proof of Proposition~\ref{prop:product}]
Starting from the joint distribution of $Y_2^n,X_1^n$ conditioned on $X_2^n,F_2$ we have that
\begin{align*}
  &P_{Y_2^n,X_1^n|X_2^n,F_2}(y_2^n,x_1^n|x_2^n,f_2) 
\\&\stackrel{\rm a)}{=} P_{X_1^n}(x_1^n)\prod_{i=1}^nP_{Y_{2i}|X_{1i},X_{2i}}(y_{2i}|x_{1i},x_{2i})
\\&\stackrel{\rm b)}{=}\prod_{i=1}^n P_{X_{1i}}(x_{1i})\prod_{i=1}^nP_{Y_{2i}|X_{1i},X_{2i}}(y_{2i}|x_{1i},x_{2i}) 
\\&\stackrel{\rm c)}{=}\prod_{i=1}^nP_{Y_{2i},X_{1i}|X_{2i}}(y_{2i},x_{1i}|x_{2i}) 
\end{align*}
where the equalities follows from:
 a) the inputs are independent and the channel is memoryless,
 b) the assumption that $X_1^n$ has a product distribution if not conditioned on $F_1$ as in \eqref{eq:osvaldo:eq}, and
 c) the inputs are independent.
By marginalizing with respect to $X_1^n$ yields
\begin{align*}
P_{Y_2^n|X_2^n,F_2}(y_2^n|x_2^n,f_2)
= \prod_{i=1}^n \sum_{x_{1i}}P_{Y_{2i},X_{1i}|X_{2i}}(y_{2i},x_{1i}|x_{2i})
= \prod_{i=1}^n P_{Y_{2i}|X_{2i}}(y_{2i}|x_{2i}),
\end{align*}
as claimed. 

\end{IEEEproof}

The main result of the section is the following upper bound:

\begin{thm}
\label{thm:outerbound}
Any achievable rate pair $(R_{1},R_{2})$ for the IC-OR  must satisfy 
\begin{subequations}
\begin{align}
R_1 &\le I(Y_{1};X_{1}|X_{2},Q),     \label{eq:R1o} && \text{(memoryless IC-OR)}\\
R_2 &\le I(Y_{2};X_{2}|Q),           \label{eq:R2o} && \text{(memoryless IC-OR)}\\
R_1+R_2 &\le  H(Y_{1}|Q)+H(Y_{2}|U_2,Q)
\notag\\&-H(T_{2}|X_{2},Q)-H(T_{1}|Q)  &&\text{(memoryless ISD-IC-OR)}
\notag\\&= I(Y_{1};X_{1},X_{2}|Q) + I(Y_{2};X_{2}|U_2,Q), \label{eq:sR1+R2 o}
\end{align}
for some input distribution that factors as 
\begin{align}
&P_{Q,X_1,X_2,U_2}(q,x_1,x_2,u_2)
=P_Q(q)P_{X_1|Q}(x_1|q)P_{X_2|Q}(x_2|q)P_{T_2|X_2}(u_2|x_2), 
\label{eq:new outer bounds input pdf}
\end{align}
\label{eq:new outer bounds}
\end{subequations}
and with $|\mathcal{Q}| \le 2$. 
We denote the region in~\eqref{eq:new outer bounds} as $\mathcal{R}_{\rm out}$.
\end{thm}

\begin{IEEEproof}[Proof of Theorem~\ref{thm:outerbound}]
By Fano's inequality
 $H(W_1|Y_1^n,F_1,F_2) \leq n \epsilon_n$ and
 $H(W_2|Y_2^n,    F_2) \leq n \epsilon_n$ for some
 $\epsilon_n \to 0$ as $n\to\infty$.
 
We begin with the $R_1$-bound (non-oblivious receiver) in~\eqref{eq:R1o}:
\begin{align*}
n(R_1-\epsilon_n)
  & \stackrel{\rm a)}{\le} I(W_1;Y_1^n,F_1,F_2)
\\&\stackrel{\rm b)}{\le} I(W_1;Y_1^n|F_1,F_2,W_2) 
\\&\stackrel{\rm c)}{\le} I(X_1^n;Y_1^n|F_1,F_2,X_2^n)
\\&\stackrel{\rm d)}{=  } H(Y_1^n|F_1,F_2,X_2^n) -\sum_{i=1}^n H(Y_{1i}|X_{1i},X_{2i})
\\&\stackrel{\rm e)}{\le} \sum_{i=1}^nH(Y_{1i}|X_{2i}) -\sum_{i=1}^n H(Y_{1i}|X_{1i},X_{2i})
\\&=  \sum_{i=1}^{n} I(X_{1i};Y_{1i}|X_{2i}),
\end{align*}
where the (in)equalities follow from:
a) Fano's inequality,
b) giving $W_2$ as side information and using the fact that  $F_1$, $F_2$, $W_1$ and $W_2$ are mutually independent, 
c) data processing $(F_i,W_i) \rightarrow X_i^n \rightarrow Y_1^n$, for $i\in[1:2]$, and
d) because the channel is memoryless, 
e) by chain rule of entropy and by ``conditioning reduces entropy''.
For the $R_2$-bound (oblivious receiver) in~\eqref{eq:R2o} we have:
\begin{align*}
n(R_2-\epsilon_n)
  &\stackrel{\rm a)}{\le} I(W_2;Y_2^n,F_2)
\\&\stackrel{\rm b)}{\le} I(W_2;Y_2^n|F_2) 
\\&\stackrel{\rm c)}{\le} I(X_2^n;Y_2^n|F_2)
\\&\stackrel{\rm d)}{=  } H(Y_2^n|F_2) -\sum_{i=1}^n H(Y_{2i}|X_{2i})
\\&\stackrel{\rm e)}{\le} \sum_{i=1}^{n}H(Y_{2i}) -\sum_{i=1}^n H(Y_{2i}|X_{2i})
\\&=  \sum_{i=1}^{n}  I(X_{2i};Y_{2i}),
\end{align*}
where the (in)equalities follow from:
a) Fano's inequality,
b) the fact that  $F_2$ and $W_2$ are independent, 
c) data processing $(F_i,W_i) \rightarrow X_i^n \rightarrow Y_1^n$, for $i\in[1:2]$, and
d) by  Proposition~\ref{prop:product}, 
e) from chain rule of entropy and ``conditioning reduces entropy''.

Next, by providing $U_2$ as side information to receiver~2 (oblivious receiver) similarly to \cite{Telatar_Tse_inject}\footnote{Random variable $U_2$ is obtained by passing $X_2$ through an auxiliary channel described by $P_{T_2|X_2}$. Intuitively, $U_2$ represents interference  caused by $X_2$ plus noise at the output $Y_1$. The idea is that providing a noisy version of $X_2$ as side information will result in a tighter bound than for example giving just $X_2$.},
where $U_2$  is jointly distributed with the inputs according to~\eqref{eq:new outer bounds input pdf},
we have: 
\begin{align*}
n(R_1+R_2-2\epsilon_n) 
&\stackrel{ \rm a)}{\le} I(X_1^n;Y_1^n|F_1,F_2)+I(X_2^n;Y_2^n,U_2^n|F_2)\\
& = H(Y_1^n|F_1,F_2)-H(Y_1^n|F_1,F_2,X_1^n)\\
&+H(U_2^n|F_2)-H(U_2^n|F_2,X_2^n)\\
&+H(Y_2^n|F_2,U_2^n)-H(Y_2^n|F_2,X_2^n,U_2^n)\\
& \stackrel{\rm b)}{=} H(Y_1^n|F_1,F_2)-H(T_2^n|F_1,F_2)\\
&+H(U_2^n|F_2)-H(U_2^n|F_2,X_2^n)\\
&+H(Y_2^n|F_2,U_2^n)-H(T_1^n)\\
& \stackrel{\rm c)}{=}H(Y_1^n|F_1,F_2)-H(T_2^n|F_1,F_2)\\
&+H(T_2^n|F_2)-H(T_2^n|F_2,X_2^n)\\
&+H(Y_2^n|F_2,U_2^n)-H(T_1^n)\\
& \stackrel{\rm d)}{=}H(Y_1^n|F_1,F_2)+H(Y_2^n|F_2,U_2^n)-H(T_2^n|X_2^n)-H(T_1^n)\\ 
& \stackrel{\rm e)}{\le} \sum_{i=1}^{n} H(Y_{1i}|F_1,F_2)+H(Y_{2i}|F_2,U_{2i}) -  H(T_{2i}|X_{2i})- H(T_{1i}),\\
& \stackrel{\rm f)}{\le} \sum_{i=1}^{n} H(Y_{1i})+H(Y_{2i}|U_{2i}) -  H(T_{2i}|X_{2i})- H(T_{1i}),
\end{align*}

where (in)equalities follow from:
a) by Fano's inequality and by giving $U_2$ as side information and by proceeding as done for the single rate bounds up to step labeled ``c)'',
b) by the injective property in~\eqref{eq:ISD def} and the independence of $(X_1^n,T_1^n)$ and $X_2^n$,
c) by definition of $U_2$ in~\eqref{eq:new outer bounds input pdf} we have $H(U_2^n|F_2)=H(T_2^n|F_2)$,
d) by independence of the messages we have  $H(T_2^n|F_1,F_2)-H(T_2^n|F_2)=0$,
e) since the channel is memoryless and thus $H(T_2^n|F_2,X_2^n)=H(T_2^n|X_2^n)=\sum_{i=1}^nH(T_{2i}|X_{2i})$ and since $H(T_1^n)=H(Y_2^n|X_2^n)$ can be single-letterized by using Proposition~\ref{prop:product}, and 
f) by conditioning reduces entropy.

The introduction of a time-sharing random variable $Q  \sim \mathsf{Unif}[1:n]$ yields the bounds in~\eqref{eq:new outer bounds}.
The Fenchel-Eggleston-Caratheodory theorem 
\cite[Chapter 14]{Csiszar:1981} guarantees that we may restrict attention to $|Q| \le 2$ without loss of optimality.

Finally, the equality in~\eqref{eq:sR1+R2 o} follows from the injective property in~\eqref{eq:ISD def}, the independence of the inputs and the memoryless property of the channel, i.e.,
\begin{align*}
& H(T_2|X_{2})
 = H(T_2|X_{1},X_{2}) 
 = H(Y_{1}|X_{1},X_{2},Q), 
\\
&H(T_1|Q)
= H(T_1|U_2,Q,X_{2}).
\end{align*}
This concludes the proof.
\end{IEEEproof}

\section{Capacity Results} 
\label{sec:ch:capacity}

In this section we prove that the outer bound in~\eqref{eq:new outer bounds} is (approximately) tight in certain regimes or for certain classes of channels. To start, we propose an achievable rate region based on a simplified Han-Kobayashi scheme \cite{H+K} in which joint decoding of the intended and interfering messages is not required at receiver~2 (the oblivious receiver) and in which every node uses an i.i.d. codebook.

\subsection{Inner Bound}

Consider an achievability scheme where encoder 1 transmits using an i.i.d. codebook, while encoder 2, corresponding to the oblivious receiver, rate-splits as in the Han and Kobayashi achievability scheme for the classical IC \cite{H+K}. It may then be shown that the following rates are achievable, 

\begin{prop}
\label{prop:innerbound}
The set of non-negative rate pairs $(R_1,R_2)$ satisfying
\begin{subequations}
\begin{align}
R_1 &\le I(Y_1;X_1|U_2,Q),\label{eq:R1i}\\
R_2 &\le I(Y_2;X_2|Q), \label{eq:R2i}\\
R_1+R_2 & \le I(Y_1;X_1,U_2|Q)+I(Y_2;X_2|U_2,Q),\label{eq:sR1+R2 i}
\end{align}
is achievable for every input distribution that factorizes as 
\begin{align}
P_{Q,X_1,X_2,U_2}=P_QP_{X_1|Q}P_{X_2|Q}P_{U_2|X_2Q},
\label{eq:inner bound input pdf}
\end{align} 
{and where $|Q|\leq 8$ from \cite{motani}.}
\label{eq:inner bound}
\end{subequations}
We denote the region in~\eqref{eq:inner bound} as $\mathcal{R}_{\rm in}$, which is achievable for any memoryless IC-OR.
\end{prop}

\begin{IEEEproof}[Proof of Proposition~\ref{prop:innerbound}]
The proof follows by setting the auxiliary r.v. $U_1$ in the Han and Kobayashi rate region in \cite[Section 6.5]{elgamalkimbook} to $U_1=\emptyset$. Note, that this modified version of the Han and Kobayashi scheme employs joint decoding (of desired and undesired messages) only at receiver 1 (the non-oblivious receiver) and hence knowledge of the codebook of transmitter 1 is not needed at receiver 2 (the oblivious receiver). 
\end{IEEEproof}

\begin{rem}
\label{rem:differences}
By comparing 
the outer bound region $\mathcal{R}_{\rm out}$ in Theorem~\ref{thm:outerbound} to
the inner bound region $\mathcal{R}_{\rm in}$  in Proposition~\ref{prop:innerbound} we notice the following differences:
1) in~\eqref{eq:new outer bounds input pdf} the side information random variable $U_2$ is distributed as $T_2$ conditioned on $X_2$, while in~\eqref{eq:inner bound input pdf} the auxiliary random variable $U_2$ can have any distribution conditioned on $X_2$;
2) the mutual information terms involving $Y_1$ have $X_2$ in the outer bound, but $U_2$ in the inner bound; and
3) the mutual information terms involving $Y_2$ are the same in both regions.
\end{rem}

\subsection{Capacity in very strong interference at the non-oblivious receiver for the general memoryless IC-OR}
In this section we show that under special channel conditions, akin to the very strong interference regime for the classical IC,
the outer bound region in Theorem~\ref{thm:outerbound} is tight.

A general memoryless IC-OR for which
\begin{align}
I(X_2;Y_2|X_1) \le I(X_2;Y_1), \quad \forall P_{X_1,X_2}=P_{X_1}P_{X_2},
\label{eq:strong interf.cond. for one Rx oblivious}
\end{align}
is said to have {\em very strong interference at the non-oblivious receiver} (receiver~1).
Intuitively, when the condition in~\eqref{eq:strong interf.cond. for one Rx oblivious} holds, the non-oblivious receiver should be able to first decode the interfering signal by treating its own signal as noise and then decode its own intended signal free of interference. This should ``de-activate'' the sum-rate bound in~\eqref{eq:sR1+R2 o}. Next we formalize this intuition.

\begin{thm}\label{thm:very strong cap memoryless}
When the condition in~\eqref{eq:strong interf.cond. for one Rx oblivious} holds the capacity region of the IC-OR is given by 
\begin{subequations}
\begin{align}
R_1 &\le I(X_1;Y_1|X_2,Q),\\
R_2 &\le I(X_2;Y_2|Q), 
\end{align}
\label{eq:outer for one Rx oblivious}
\end{subequations}
taken over the union of all input distributions that factor as $P_{Q,X_1,X_2}=P_{Q}P_{X_1|Q}P_{X_2|Q}$
{and where $|Q|\leq 2$.}
\end{thm}
\begin{IEEEproof}[Proof of Theorem~\ref{thm:very strong cap memoryless}]
By dropping the sum-rate outer bound in~\eqref{eq:sR1+R2 o} we see that the region in~\eqref{eq:outer for one Rx oblivious} is an outer bound for a general memoryless IC-OR. By setting $U_2=X_2$ in the achievable region in~\eqref{eq:inner bound}, the region 
\begin{subequations}
\begin{align}
R_1 &\le I(X_1;Y_1|X_2,Q),  \label{eq:eq1:strong:proof}\\
R_2 &\le I(X_2;Y_2|Q),      \label{eq:eq2:strong:proof}\\
R_1+R_2 &\le I(X_1,X_2;Y_1|Q), \label{eq:sumrate:strong:proof}
\end{align}
\label{eq:inner for one Rx oblivious}
\end{subequations}
taken over the union of all $P_{Q,X_1,X_2}=P_{Q}P_{X_1|Q}P_{X_2|Q}$, is achievable.
We see that the single rate bounds in~\eqref{eq:inner for one Rx oblivious} match the upper bounds in~\eqref{eq:outer for one Rx oblivious}. We next intend to show that when the condition in~\eqref{eq:strong interf.cond. for one Rx oblivious} holds, the sum-rate bound in~\eqref{eq:sumrate:strong:proof} is redundant.
By summing \eqref{eq:eq1:strong:proof} and \eqref{eq:eq2:strong:proof}
\begin{align*}
R_1+R_2
  &\le I(X_1;Y_1|X_2,Q) + I(X_2;Y_2|Q)     
\\&\stackrel{\rm a)}{\le} I(X_1;Y_1|X_2,Q) + I(X_2;Y_2,X_1|Q) 
\\&\stackrel{\rm b)}{=}   I(X_1;Y_1|X_2,Q) + I(X_2;Y_2|X_1,Q) 
\\&\stackrel{\rm c)}{\le} I(X_1;Y_1|X_2,Q) + I(X_2;Y_1|Q)     
\\&=   I(X_1,X_2;Y_1|Q) = \text{eq.\eqref{eq:sumrate:strong:proof}},                    
\end{align*}
where
in a) we loosened the achievable sum-rate by adding $X_1$ as ``side information'' to receiver~2,
in b) we used the independence of the inputs, and
in c) the condition in~\eqref{eq:strong interf.cond. for one Rx oblivious}.
Therefore, the sum-rate bound in~\eqref{eq:sumrate:strong:proof} can be dropped without affecting the achievable rate region. This shows that the outer bound in~\eqref{eq:outer for one Rx oblivious} is achievable thereby proving the claimed capacity result.
\end{IEEEproof}

\begin{rem}
For the classical IC, the very strong interference regime is defined as
\begin{align*}
I(X_1;Y_1|X_2) \le I(X_1;Y_2),\\
I(X_2;Y_2|X_1) \le I(X_2;Y_1),
\end{align*}
for all product input distributions; under these pair of conditions capacity can be shown.
For the IC-OR, the very strong interference constraint at receiver~2 (oblivious receiver) is not needed in order to show capacity. Therefore, the very strong interference condition for the IC-OR is less stringent than that for the classical IC.  We believe this is so because the oblivious receiver (receiver~2) cannot decode the message of user~1 as per the modeling assumption. Indeed, we feel that the ``lack of codebook knowledge'' as originally proposed in \cite{sand_decentr_proces} actually models the inability of a receiver to jointly decode its message along with unintended ones, as the mapping between the messages and codewords is not known.
\end{rem}

\subsection{Capacity to within a Constant Gap for the ISD-IC-OR}
We now show that $\mathcal{R}_{\rm in}$ in Proposition~\ref{prop:innerbound} lies to within a  gap of the outer bound $\mathcal{R}_{\rm out}$ in Theorem~\ref{thm:outerbound} for the general ISD-IC-OR. We have

\begin{thm}
\label{thm:gap}
For the ISD-IC-OR, if $(R_1,R_2) \in \mathcal{R}_{\rm out}$ 
then 
$([R_1-I(X_2;T_2|U_2,Q)]^+,R_2) \in \mathcal{R}_{\rm in}.$
\end{thm}
\begin{IEEEproof}[Proof of Theorem~\ref{thm:gap}]
The proof is as in \cite{Telatar_Tse_inject}.
First, we define a new outer bound region $\bar{\mathcal{R}}_{\rm out}$ by  replacing $X_2$ with $U_2$ 
in all \emph{positive} entropy terms of region $\mathcal{R}_{\rm out}$,  which is permitted as  $H(Y_2|X_2) \le H(Y_2|U_2)$ by the data processing inequality. 
We conclude that $  \mathcal{R}_{\rm out} \subseteq\bar{\mathcal{R}}_{\rm out}$.
We next compare $\bar{\mathcal{R}}_{\rm out}$ and  $\mathcal{R}_{\rm in}$ term by term (we only need to compare the mutual informations invoking $Y_1$ as those involving $Y_2$ are the same in both bounds, see Remark~\ref{rem:differences}, thus implying a zero gap for rate $R_2$): 
the difference is that $\bar{\mathcal{R}}_{\rm out}$ has $-H(Y_{1}|X_1,X_2)$ where $\mathcal{R}_{\rm in}$ has $-H(T_2|U_2,Q)$; 

thus the gap is 
\[
-H(Y_{1}|X_1,X_2) + H(T_2|U_2,Q)
= -H(T_2|X_2)+H(T_2|U_2,Q)
= I(X_2;T_2|U_2,Q).
\]
This concludes the proof.
\end{IEEEproof}

\begin{rem}
Note that 
\[
I(X_2;T_2|U_2,Q) 
= H(T_2|U_2,Q)-H(T_2|X_2)
\leq H(T_2)-H(T_2|X_2)
\leq \max_{p_{X_2}} I(T_2;X_2),
\] 
so the gap is finite / constant for all channel $P_{T_2|X_2}$ with finite capacity.
\end{rem}

We next give an example of constant gap characterization in Section~\ref{sec:ch:Gaussian} after having discussed in Section~\ref{sec:get no gap} a special class of ISD-IC-OR for which the gap to capacity is zero.

\subsection{Exact Capacity for the Injective Fully Deterministic IC-OR}
\label{sec:get no gap}

We now specialize Theorem \ref{thm:gap} to the class of injective {\it fully deterministic} ICs \cite{elgamal_det_IC}.
For this class of channels the mappings $T_1$ and $T_2$ in~\eqref{eq:ISD def} are deterministic functions of $X_1$ and $X_2$, respectively. We have
\begin{cor}
\label{cor:cap:Deterministic}
For the injective fully deterministic IC-OR the outer bound in Theorem~\ref{thm:outerbound} is tight.
\end{cor}
\begin{IEEEproof}[Proof of Corollary~\ref{cor:cap:Deterministic}]
The injective fully deterministic IC-OR has $T_2 = U_2$ and therefore $I(X_2;T_2|U_2,Q)=0$ in Theorem~\ref{thm:gap}.
\end{IEEEproof}

As an application of Corollary~\ref{cor:cap:Deterministic} we consider next the Linear Deterministic Approximation (LDA) of the Gaussian IC-OR at high SNR, whose classical counterpart (where all codebooks are known) was first proposed in \cite{AvestDeterministic}. 
The LDA-IC-OR has input/output relationship
\begin{subequations}
\begin{align}
&Y_1=\mathbf{S}^{q-n_{11}} X_1+\mathbf{S}^{q-n_{12}} X_2, \quad T_2=\mathbf{S}^{q-n_{12}} X_2, \\
&Y_2=\mathbf{S}^{q-n_{21}} X_1+\mathbf{S}^{q-n_{22}} X_2, \quad T_1=\mathbf{S}^{q-n_{21}} X_1,
\end{align}
\label{eq:LDA model}
\end{subequations}
where 
inputs and outputs are binary-valued vectors of length $q$,
$\mathbf{S}$ is the $q \times q$ shift matrix \cite{AvestDeterministic},
$(n_{11},n_{12},n_{21},n_{22})$ are non-negative integers and
$q:=\max \{n_{11},n_{12},n_{21},n_{22}\}$.
Summations and multiplications are bit-wise over the binary field.

For simplicity, we next evaluate the {\em symmetric} sum-capacity of the LDA-IC-OR.
The symmetric LDA-IC-OR has parameters 
$n_{11}=n_{22}=n_{\rm S}$ and $n_{12}=n_{21}=n_{\rm I} : = n_{\rm S} \ \alpha$ for some non-negative $\alpha$.
The maximum symmetric rate, or
symmetric sum-capacity normalized by the sum-capacity of an interference-free channel, 
is defined as
\begin{align}
d(\alpha) &:= \frac{\max\{R_1+R_2\}}{2 \ n_{\rm S}},
\end{align}
where the maximization is over all achievable rate pairs $(R_1,R_2)$ satisfying Theorem \ref{thm:outerbound}, which is the capacity region by Corollary~\ref{cor:cap:Deterministic}.
Since we may provide the oblivious receiver in the LDA-IC-OR with the additional codebook index so as to obtain the classical LDA-IC with full codebook knowledge, we immediately have
\begin{align}
d(\alpha) &\leq d^{\rm(W)}(\alpha) 
=\min \left(1, \max \left(\frac{\alpha}{2},1-\frac{\alpha}{2} \right), \max \left(\alpha, 1-\alpha \right) \right) , \label{eq:Wcurve}
\end{align}
where $d^{\rm(W)}(\alpha)$, the so-called W-curve \cite{etkin_tse_wang}, is the maximum symmetric rate of the classical LDA-IC.
In~\cite{bresler_tse} it was shown that i.i.d. $\mathsf{Bernoulli}(1/2)$ input bits in the Han and Kobayashi region yield $d^{\rm(W)}(\alpha)$.

\begin{figure}
\center
\includegraphics[width=9cm]{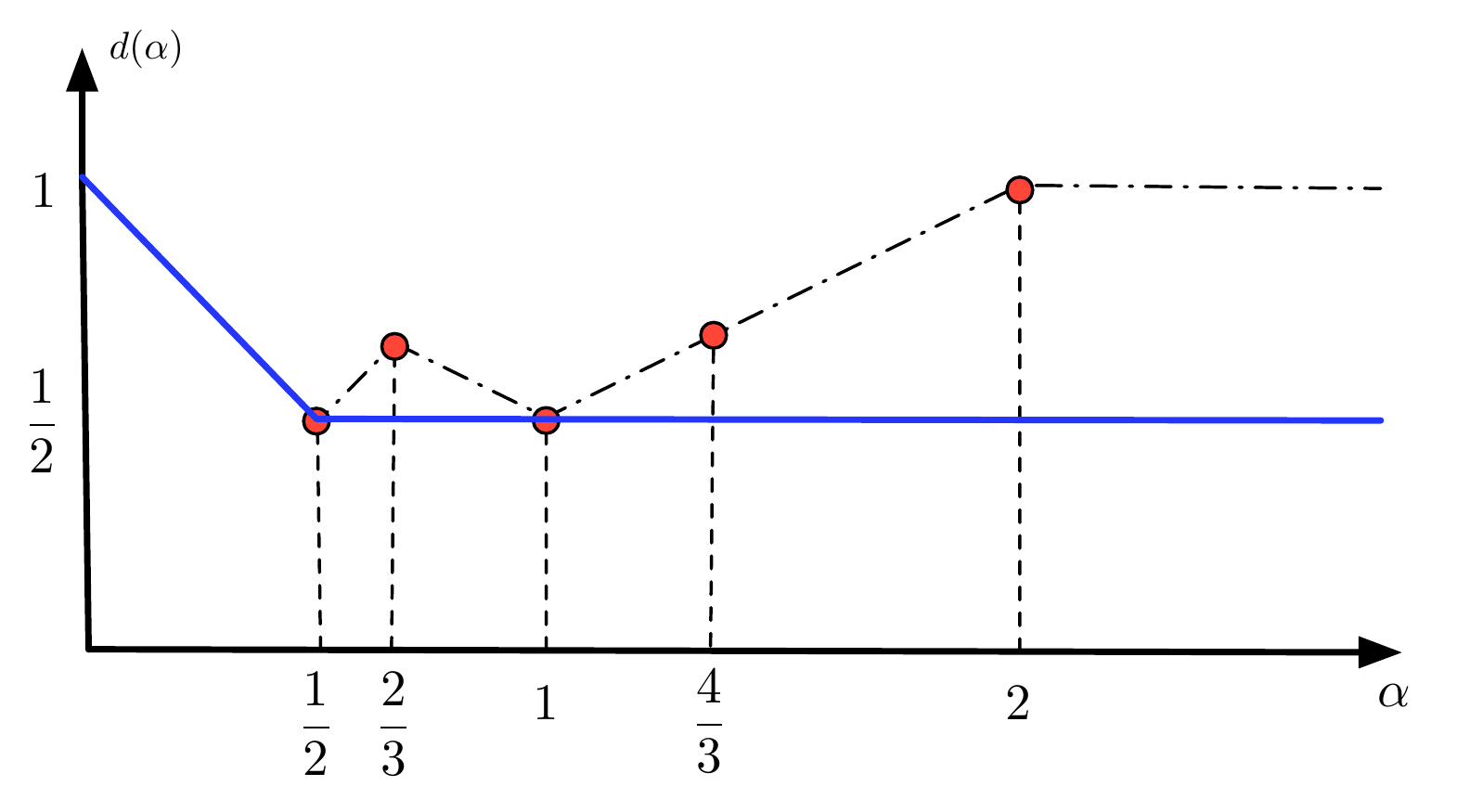}
\caption{%
The normalized sum-capacity, or maximum symmetric rate, for the classical LDA-IC (dash-dotted black line).
Normalized sum-rates achieved by the input distributions in Table~\ref{tab:tab} (red dots) for the LDA-IC-OR.
The normalized sum-rate achieved by i.i.d. $\mathsf{Bernoulli}(1/2)$ inputs and $|Q|=1$ (solid blue line) in the capacity region in Theorem \ref{thm:outerbound} for the LDA-IC-OR.}
\label{fig:LDIC}
\end{figure}

Although Theorem \ref{thm:outerbound} gives the exact capacity region of the LDA-IC-OR, it is not immediately clear which input distribution achieves the maximum symmetric rate. Instead of analytically deriving the sum-capacity, we proceeded to numerically evaluate Theorem \ref{thm:outerbound} for $|Q|=1$, which is not necessarily optimal.
We observe the surprising result that even with $|Q|=1$  i.e., without time sharing, some of the points on the normalized sum-capacity of the LDA-IC-OR are equal to $d^{\rm(W)}(\alpha)$, see Fig.~\ref{fig:LDIC} and Table~\ref{tab:tab}. Although we lack a formal proof that we can achieve the whole W-curve with a non i.i.d. $\mathsf{Bernoulli}(1/2)$ input we do, however, conjecture that it is indeed possible with the scheme in Proposition~\ref{prop:innerbound}. If true, this  implies that partial codebook knowledge at one receiver does not impact the sum-rate of the symmetric LDA-IC-OR at these points. This is quite unexpected, especially in the strong interference regime ($\alpha \geq 1$) where the optimal strategy for the classical LDA-IC is to jointly decode the interfering message along with the intended message---a strategy that seems to be precluded by the lack of codebook knowledge at one receiver. This might suggest a more general principle: there is no loss of optimality in lack of codebook knowledge as long as the oblivious receiver can remove the interfering codeword, regardless of whether or not it can decode the message carried by this codeword.

Another interesting observation is that  i.i.d. $\mathsf{Bernoulli}(1/2)$ input bits may no longer be optimal (though we do  not show their strict sub-optimality). 
In Table~\ref{tab:tab} we report, for some values of $\alpha$ and $n_{\rm S},n_{\rm I}$, the input distributions to be used in $\mathcal{R}_{\rm out}$ in Theorem \ref{thm:outerbound}. We notice that, at least when evaluating the region in Theorem \ref{thm:outerbound} for $|Q|=1$ only, that the region exhausting inputs are now {\em correlated}. 
For example, Table~\ref{tab:tab} shows that, for $\alpha=4/3$  the inputs $X_1$ and $X_2$ are binary vectors of length $\log(16)=4$ bits; out of the  $16$ different possible bit sequences, only $4$ are actually used at each transmitter with strictly positive probability to achieve $d^{\rm(W)}(4/3) = 4/6$.
By using i.i.d. $\mathsf{Bernoulli}(1/2)$ input bits in Theorem \ref{thm:outerbound} for $|Q|=1$ we would obtain a normalized sum-rate of $1/2=3/6$, the same as achieved by time division \cite{bresler_tse}.

\begin{table}
\center
\caption{LDA-IC-OR: examples of sum-rate optimal input distributions for the capacity region in Theorem \ref{thm:outerbound}.}
\label{tab:tab}
\label{Table:probabilities}
\begin{tabular}{|l|l|}
\hline
$\alpha$, $(n_{ \rm S},n_{\rm I})$ & Probability mass function with $|Q|=1$\\
\hline
$\frac{1}{2}$, $(2,1)$ & $P_{X_1}=[0.5, 0, 0.5,  0]$ \\
              & $P_{X_2}=[0, 0.5, 0,  0.5]$\\
\hline
$\frac{2}{3}$, $(3,2)$ & $ {P_{X_1}=[0,0,0.25,0.25,0,0,0.25,0.25]}$\\
              & $ {P_{X_2}=[0,0,0.25,0.25,0,0,0.25,0.25]}$\\
\hline
$1$, $(2,2)$ & $P_{X_1}=[0,0,0.5,0.5]$\\
    & $P_{X_2}=[0,0.5,0,0.5]$\\
\hline
$\frac{4}{3}$, (3,4) & ${P_{X_1}=[0, 0, 0, 0,    0, 0.25, 0, 0.25, 0, 0, 0,  0, 0,0.25, 0, 0.25]}$ \\
              & ${P_{X_2}=[0, 0, 0, 0.25, 0, 0.25, 0,    0, 0, 0, 0, 0,0,  0.25, 0, 0.25]}$\\
\hline
$2$, $(2,1)$ & $P_{X_1}=[0,0.5,0,0.5]$\\
  & $P_{X_2}=[0,0.5,0,0.5]$ \\
\hline
\end{tabular}
\end{table}

Also,  i.i.d. $\mathsf{Bernoulli}(1/2)$ inputs  in the LDA model usually are translated to i.i.d. Gaussian inputs in the Gaussian noise model. This intuition is  reinforced, in the next section, by showing that i.i.d. Gaussian are also suboptimal for the Gaussian noise model for $|Q|=1$. Also, the fact that there exist other, non i.i.d $\mathsf{Bernoulli}(1/2)$,  input distributions that are capacity achieving for the LDA stimulates search for non-Gaussian inputs that might be capacity achieving for a Gaussian noise channel. In fact the rest of the paper tries to use intuition gained in this section to construct non-Gaussian inputs that will be capacity or constant gap capacity approaching. 

\subsection{The Gaussian Noise IC-OR} 
\label{sec:ch:Gaussian} 
We now consider the practically relevant real-valued single-antenna power-constrained Gaussian noise channel, whose input/output relationship is
\begin{subequations}
\begin{align}
&Y_1=h_{11}X_1+h_{12}X_2+Z_1=h_{11}X_1+T_2, \quad T_2=h_{12}X_2+Z_1, \\
&Y_2=h_{21}X_1+h_{22}X_2+Z_2=h_{22}X_2+T_1, \quad T_1=h_{21}X_1+Z_2, 
\end{align}
\label{ch:gauss}
\end{subequations}
where $h_{ij}$  are the real-valued  channel coefficients for $(i,j) \in [1:2]^2$ assumed constant and known to all nodes,
the input  $X_i \in \mathbb{R}$  is subject to per block power constraints $\frac{1}{n}\sum_{i=1}^n X_i^2 \le 1, \ i\in[1:2]$, and 
the noise $Z_i, \ i\in[1:2],$ is a unit-variance zero-mean Gaussian r.v..

By specializing the result of Theorem~\ref{thm:gap} to the G-IC-OR we may show the following:
\begin{cor}
\label{gap:const:gaus}
For the G-IC-OR the gap is at most $1/2$~bit per channel use.
\end{cor}
\begin{IEEEproof}[Proof of Corollary~\ref{gap:const:gaus}]
For the G-IC-OR $T_2=h_{12}X_2+Z_1$,  and 
thus we set $U_2$ in Theorem~\ref{thm:outerbound} to $U_2=h_{12}X_2+Z_1^{*}$,
where $Z_1\sim Z_1^{*}$ and mutually independent.  We thus have
\begin{align*}
I(X_2;T_2|U_2,Q)
  &=   h(T_2|U_2,Q)-h(Z_2)
\\&\le h(T_2-U_2)-h(Z_1)
\\&=   h(Z_1-Z_1^{*})-h(Z_1) 
  = \frac{1}{2}\log(2),
\end{align*}
as claimed.
\end{IEEEproof}

In the classical G-IC with full codebook knowledge, Gaussian inputs exhaust known outer bounds, which are achievable to within $1/2$~bit per channel use~\cite{etkin_tse_wang}.
From the rate expression in Theorem~\ref{thm:outerbound} it is not clear 
whether Gaussian inputs are optimal for $\mathcal{R}_{\rm out}$.
The following discussion shows that in general the answer is in the negative.

For simplicity we focus on the achievable generalized Degrees of Freedom (gDoF) for the symmetric G-IC-OR.
The symmetric G-IC-OR has $|h_{11}|^2=|h_{22}|^2={\snr}$ and $|h_{12}|^2=|h_{21}|^2={\inr}$, with ${\inr} = {\snr}^\alpha$ for some non-negative $\alpha$. The sum-gDoF is defined as
\begin{align} 
d(\alpha)
:= \lim_{\snr\to+\infty } \frac{\max\{R_1+R_2\}}{2 \cdot \frac{1}{2}\log(1+{\snr})},
\label{eq:gDoF:def} 
\end{align}
where the maximization is over all possible achievable rate pairs.
By using the classical G-IC as a trivial upper bound, we have $d(\alpha)\leq d^{\rm(W)}(\alpha)$ where $d^{\rm(W)}(\alpha)$ is given in~\eqref{eq:Wcurve}.

By evaluating Theorem \ref{thm:outerbound} for independent Gaussian inputs and $|Q|=1$ (which we do {\it not} claim to be optimal, but which gives us an achievable rate up to $1/2$~bit) we obtain
\begin{align*}
(R_1+R_2)^{\rm(GG)} =   \min \Big\{   
  &\mug\left({\snr}\right)+\mug\left(\frac{{\snr}}{1+{\inr}}\right), 
\\&\mug\left(\frac{{\snr}}{{\inr}+1}\right)+\mug\left({\inr}+\frac{{\snr}}{1+{\inr}}\right) \Big\},
\\&\hspace*{-2cm}\Longleftrightarrow \quad
d^{\rm(GG)}(\alpha)= \frac{1}{2}+\left[\frac{1}{2}-\alpha\right]^{+},
\end{align*}
the superscript ``GG'' indicates that both transmitters use a Gaussian input. 

For future reference, with Time Division (TD) and Gaussian codebooks we can achieve
\begin{align*}
(R_1+R_2)^{\rm(TD)} =  \frac{1}{2} \log\left(1+ 2\ {\snr}\right)
\quad \Longleftrightarrow \quad
d^{\rm(TD)}(\alpha)
= \frac{1}{2}.
\end{align*}

We plot the achievable gDoF vs. $\alpha$ in Fig.~\ref{fig:LDIC}, together with the gDoF of the classical G-IC given by $d^{\rm(W)}(\alpha)$~\cite{etkin_tse_wang}. 
We note that Gaussian inputs are indeed optimal for $0 \le \alpha \le 1/2$, i.e., $d^{\rm(GG)}(\alpha)=d^{\rm(W)}(\alpha)$, where interference is treated as noise even for the classical G-IC (which is also achievable by the G-IC-OR). For $\alpha > 1/2$ we have $d^{\rm(GG)}(\alpha)=d^{\rm(TD)}(\alpha)$, that is, Gaussian inputs achieve the same rates as time division. 
Interestingly, Gaussian inputs are sub-optimal in our achievable region in general as we show next.

\begin{figure}
\center
\includegraphics[width=9cm]{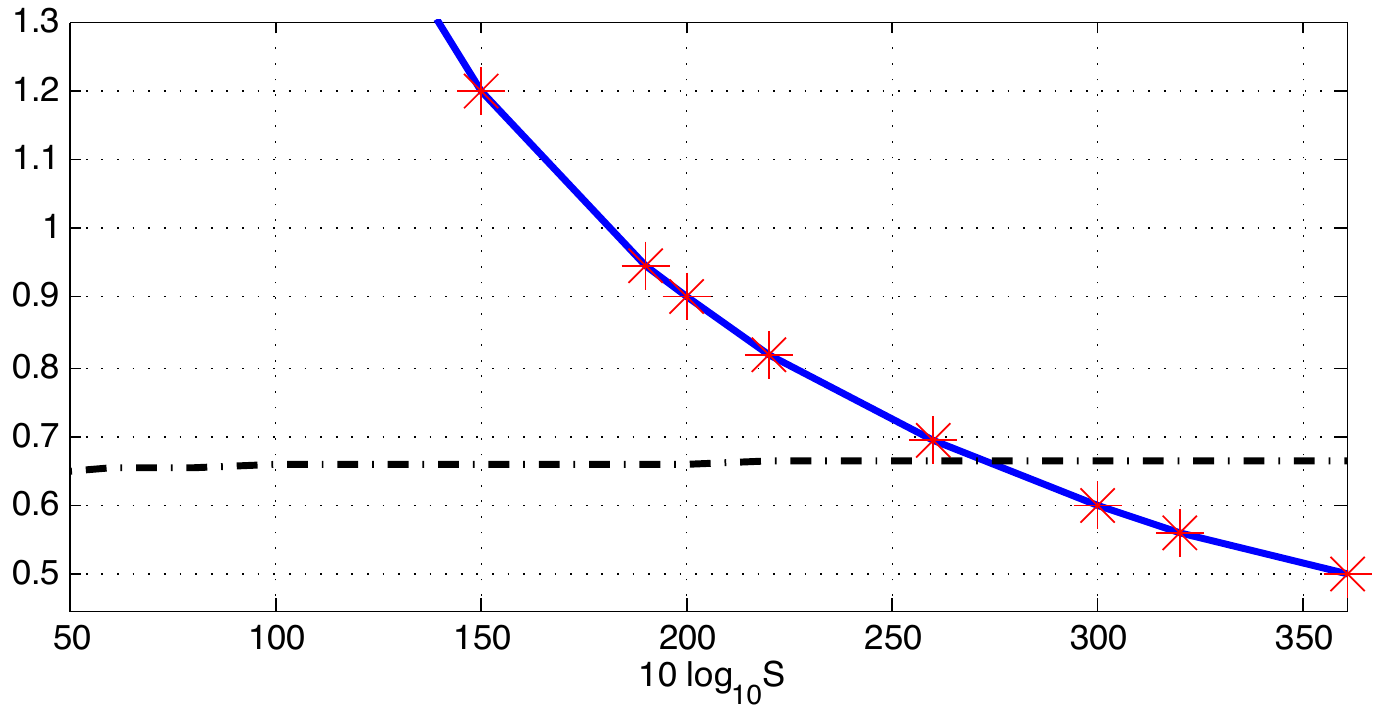}
\caption{Achievable normalized sum-rate for the symmetric G-IC-OR with $\alpha=4/3$ vs $\snr$ in dB. Legend:
time division in solid blue line; 
Gaussian inputs at both transmitters in red stars; 
$X_1$ is a uniform PAM with $N=\lfloor \snr^{\frac{1}{6}} \rfloor$ points and $X_2$ is Gaussian in dash-dotted black line.}  
\label{fig:alpha4/3}
\end{figure}

Consider $\alpha=4/3$. With Gaussian inputs 
we only achieve $d^{\rm(GG)}(4/3)=d^{\rm(TD)}(4/3)=1/2$. Notice the similarity with the LDA-IC-OR: the input distribution that is optimal for the non-oblivious IC performs as time division for the G-IC-OR. Inspired by the LDA-IC-OR we explore now the possibility of using non-Gaussian inputs. 
By following \cite[Section VI.A]{sand_decentr_proces}, which demonstrated that binary signaling outperforms Gaussian signaling for a fixed finite $\snr$, we consider a uniform PAM constellation with $N$ points.
Fig. \ref{fig:alpha4/3} shows the achievable normalized sum-rate $\frac{R_1+R_2}{2 \cdot \frac{1}{2}\log(1+\snr)}$ as a function of $\snr$ for the case where $X_1$ (the input of the non-oblivious pair) is a PAM constellation with $N=\left\lfloor \snr^{{1}/{6}} \right\rfloor$ points and $X_2$ (the input of the oblivious pair) is Gaussian; we refer to the achievable gDoF of this inputs as  $d^{\rm(DG)}(\alpha)$. Notice that the number of points in the discrete input is a function of  $\snr$.
We also report the achievable normalized sum-rate with time division and Gaussian inputs. Fig. \ref{fig:alpha4/3} shows that, for sufficiently large $\snr$, using a discrete input outperforms time division; moreover, for the range of simulated $\snr$, it seems that the proposed discrete input achieves a gDoF of $d^{\rm(DG)}(\alpha)=\alpha/2=4/6$ as for the classical G-IC with full codebook knowledge. 
In the sections that follow we analytically show that using discrete input (or mixed) at the non-oblivious transmitter indeed achieves the full gDoF and symmetric capacity region to within a constant gap.

\section{Discrete Inputs: main tool}
\label{sec:MainTool}
In this section we review the lower bound of \cite{PAMozarow} on the mutual information achievable by a PAM input on a point-to-point power-constrained Gaussian noise channel that will serve as the main tool in evaluating our inner bound for the G-IC-OR in Proposition~\ref{prop:innerbound}.
The bound is as follows:

\begin{thm} 
\label{thm:lowbound}
Let $X_D \sim \mathsf{PAM}(N)$ and let $Z_G\sim \mathcal{N}(0,1)$ and $\snr$ be a non-negative constant. 
Then
\begin{align}
 \left[ \mug\left(\min\Big(N^2-1, \snr \Big)\right) -\frac{1}{2}\log\left(\frac{\pi\eu}{3}\right)\right]^+ 
 =: \mud\left(N, \snr \right)
 \label{eq:lower Xd}
 \\ 
 \leq I(X_D; \sqrt{\snr} \ X_D+Z_G) 
 \leq  \mug\left(\min\Big(N^2-1, \snr \Big)\right).
 \label{eq:upper Xd}
\end{align}
\end{thm}
\begin{IEEEproof}[Proof of Theorem~\ref{thm:lowbound}]
The upper bound in~\eqref{eq:upper Xd} follows from the well known facts that ``Gaussian maximizes the differential entropy for a given second moment constraint'' and that ``a uniform input maximizes the entropy of a discrete random variable'' \cite{elgamalkimbook}.
Let now $x_{\min}:=\min(N^2-1, \snr)$ and $x_{\max}:=\max(N^2-1, \snr)$. We have
\begin{align*}
  &I(X_D; \sqrt{\snr}X_D+Z_G) 
\\&\stackrel{\text{from \cite[Part b]{PAMozarow}}}{\geq} 
   \frac{1}{2}\log\left(1+(N^2-1)\right)-\frac{1}{2}\log\left(1+ \frac{N^2-1}{1+\snr} \right)
  -\frac{1}{2}\log\left(\frac{\pi\eu}{6}\right)
\\&=\mug\left(x_{\min}\right)+\mug\left(x_{\max}\right)-\mug\left(x_{\min}+x_{\max}\right)
  -\frac{1}{2}\log\left(\frac{\pi\eu}{6}\right)
\\&=\mug\left(x_{\min}\right)-\mug\left(\frac{x_{\min}}{1+x_{\max}}\right)
  -\frac{1}{2}\log\left(\frac{\pi\eu}{6}\right)
\\&\geq \mug\left(x_{\min}\right)
  -\frac{1}{2}\log\left(\frac{\pi\eu}{3}\right),
\end{align*}
since $\frac{x_{\min}}{1+x_{\max}}\in[0,1]$. This, combined with non-negativity of mutual information, gives the lower bound in~\eqref{eq:lower Xd}.
\end{IEEEproof}

\begin{rem}\label{rem:on nature of gap}
The upper and lower bounds in Theorem~\ref{thm:lowbound} are to within $\frac{1}{2}\log\left(\frac{\pi\eu}{3}\right)$ bits of one another. We shall refer to the quantity $\frac{1}{2}\log\left(\frac{\pi\eu}{3}\right)$ as the ``shaping loss'' due to the use of a one-dimensional lattice constellation on the power-constrained point-to-point Gaussian channel. 
Note that what is known as ``shaping gain'' of a one-dimensional lattice constellation in the literature is $\frac{1}{2}\log\left(\frac{\pi\eu}{6}\right)$ \cite{Forney:shaping}; what we call here ``shaping loss'' has an extra $\frac{1}{2}\log(2)$ due to the term $\mug\left(\frac{x_{\min}}{1+x_{\max}}\right)$; we refer to the sum of these two contributions as ``shaping loss'' because it is purely due to the one-dimensional lattice (``shaping'' part) and it causes a reduction in rate compared to the upper bound (``loss'' part).

If we could choose $N^2-1=\snr \Longleftrightarrow N=\sqrt{1+\snr}$ then we could claim that a PAM input is optimal (i.e., achieves the capacity of the point-to-point power-constrained Gaussian noise channel) to within $\mathsf{gap} \leq \frac{1}{2}\log\left(\frac{\pi\eu}{3}\right)$  bits per channel use, where the gap would be completely due to the shaping loss.

Unfortunately, $N$ is constrained to be an integer. If for $N$ we choose the closest integer to $\sqrt{1+\snr}$, that is, $N=\left \lfloor \sqrt{1+\snr} \right\rfloor =: \points(\snr)$, then we incur a further 1~bit ``integer penalty'', by which we mean that the difference between the point-to-point Gaussian channel capacity and the lower bound on the achievable rate with a PAM in~\eqref{eq:lower Xd} is upper bounded as
\begin{align}
\mathsf{gap} 
  &\leq \mug\left(\snr\right) - \mud\left( \points(\snr), \snr \right) \notag
\\&\leq \frac{1}{2}\log\left(\frac{\pi\eu}{3}\right)
       +\frac{1}{2}\log^+\left(\frac{1+\snr}{\left \lfloor \sqrt{1+\snr} \right\rfloor^2}\right) \notag
\\&\leq \underbrace{\frac{1}{2}\log\left(\frac{\pi\eu}{3}\right)}_{\text{shaping loss}}
       +\underbrace{\frac{1}{2}\log(4)}_{\text{integer penalty}}
           =\frac{1}{2}\log\left(\frac{4\pi\eu}{3}\right),
\label{eq:shaping and integer losses}
\end{align}
where the largest integer penalty is attained for $1+\snr = 2^2 -\epsilon, \ 0<\epsilon \ll 1$, for which $\lfloor \sqrt{1+\snr} \rfloor^2 = (2-1)^2=1$. 

\end{rem}

\section{Achievable Regions for the G-IC-OR}
\label{sec:achScheme}

We now analyze the G-IC-OR by using Theorem~\ref{thm:lowbound} (i.e., bounds on the mutual information achievable by a PAM input on a point-to-point power-constrained Gaussian noise channel) and the insight on the nature of the gap due to a PAM input from Remark~\ref{rem:on nature of gap}.
We first present a scheme (an achievable rate region evaluated using a mixed input) that will prove to be useful in strong and very strong interference, and then present a more involved scheme that will be useful in the somewhat trickier weak and moderate interference regimes. 
Although the second scheme includes the first as a special case, we start with a simpler scheme to highlight the important steps of the derivation without getting caught up in excessive technical details.

\subsection{Achievable Scheme I}
We first derive an achievable rate region from Proposition~\ref{prop:innerbound} with inputs 
\begin{subequations}
\begin{align} 
\text{Scheme~I:} \quad
&X_{1D} \sim \mathsf{PAM}\left(N\right), \ N\in\mathbb{N}, \ \text{independent of}
\\
&X_{2G} \sim {\mathcal{N}(0,1)},
\\
&X_1 = X_{1D}, \ X_2 = X_{2G}, 
\\
&U_2 = X_2, \ Q=\emptyset.
\end{align}
\label{eq:inputs scheme1}
\end{subequations}

which we will show in the next sections to be gDoF optimal and to within a constant gap of the symmetric capacity of the classical G-IC in the strong and very strong interference regimes. Such results may not be shown by using i.i.d. Gaussian inputs in the same achievable scheme in Proposition~\ref{prop:innerbound}. The achievable region is derived for a general G-IC-OR and later on specialized to the symmetric case.
\begin{thm}
\label{thm:sec:achScheme G-IC-OR}
For the G-IC-OR the following rate region is achievable by the input in \eqref{eq:inputs scheme1}
\begin{subequations}
\begin{align}
R_1 &\leq \mud\left(N, |h_{11}|^2\right), 
\label{eq:ach.reg. Gaussian ICOR U2=X2 R1}
\\
R_2 & \leq\mud\left(N, \frac{|h_{21}|^2}{1+|h_{22}|^2}\right) + \mug\left(|h_{22}|^2\right)
\notag\\&-\mug\left( \min \left(N^2-1,|h_{21}|^2 \right) \right), 
\label{eq:ach.reg. Gaussian ICOR U2=X2 R2}
\\
R_1+R_2 &\leq\mud\left(N, \frac{|h_{11}|^2}{1+|h_{12}|^2}\right) + \mug\left(|h_{12}|^2\right).
\label{eq:ach.reg. Gaussian ICOR U2=X2 R1+R2}
\end{align}
\label{eq:ach.reg. Gaussian ICOR U2=X2}
\end{subequations}
\end{thm}
\begin{IEEEproof}[Proof of Theorem~\ref{thm:sec:achScheme G-IC-OR}]
We proceed to evaluate the rate region in Proposition~\ref{prop:innerbound} with the inputs in~\eqref{eq:inputs scheme1}, that is,  the achievable region in~\eqref{eq:inner for one Rx oblivious} with $|Q|=1$.

The rate of the user~1 is bounded by $R_1\leq I(X_1;Y_1|X_2)=I(X_{1D}; h_{11} X_{1D} + Z_1)$, where $I(X_{1D}; h_{11} X_{1D} + Z_1)$ can be further lower bounded by using~\eqref{eq:lower Xd} from Theorem~\ref{thm:lowbound} with $\snr=|h_{11}|^2$; by doing so we obtain the bound in~\eqref{eq:ach.reg. Gaussian ICOR U2=X2 R1}.

The rate of the user~2 is bounded by  
\begin{align*}
 R_2 & \leq I(X_2;Y_2) 
\\&=h(h_{21}X_{1D}+\underbrace{h_{22}X_{2G}+Z_2}_{\sim\mathcal{N}(0,1+|h_{22}|^2)}) -h(h_{21}X_{1D}+Z_2) 
\\&=\underbrace{\left(h\left(\frac{h_{21}}{\sqrt{1+|h_{22}|^2}}X_{1D}+Z_2\right)-h(Z_2)\right)}_{\geq \mud\left(N, \frac{|h_{21}|^2}{1+|h_{22}|^2}\right) \ \text{from~\eqref{eq:lower Xd}}}
   +\frac{1}{2}\log(1+|h_{22}|^2) 
\\&-\underbrace{\Big(h\left(h_{21}X_{1D}+Z_2\right)-h(Z_2)\Big)}_{\leq \mug\left( \min \left(N^2-1,|h_{21}|^2 \right) \right)  \ \text{from~\eqref{eq:upper Xd}}},
\end{align*}
from which we conclude that the achievable rate for user~2 is lower bounded as in~\eqref{eq:ach.reg. Gaussian ICOR U2=X2 R2}.

The sum-rate is bounded by $R_1+R_2 \leq I(X_1,X_2;Y_1)=I(X_1;Y_1)+I(X_2;Y_1|X_1)$, where 
 $I(X_1;Y_1)$ can be lower bounded by means of Theorem~\ref{thm:lowbound}
with $\snr=\frac{|h_{11}|^2}{1+|h_{12}|^2}$  and 
where $I(X_2;Y_1|X_1)=I(X_{2G}; h_{12}X_{2G}+Z_1)=\mug(|h_{12}|^2)$; by combining the two terms we obtain the bound in~\eqref{eq:ach.reg. Gaussian ICOR U2=X2 R1+R2}.

\end{IEEEproof}

\subsection{Achievable Scheme II}
The input in \eqref{eq:inputs scheme1} might not be optimal in general and may be generalized as follows.
Consider the rate region in Proposition~\ref{prop:innerbound} with inputs
\begin{subequations}
\begin{align}
\text{Scheme~II:} \quad
   &\text{$X_{1D},X_{1G},X_{2Gc},X_{2Gp}$ independent and distributed as}
\\ &X_{1D}\sim \mathsf{PAM}\left(N\right), \ N\in\mathbb{N},
\\ &\text{all the others are $\mathcal{N}(0,1)$}, 
\\ &X_1 = \sqrt{1-\delta_1} X_{1D}  + \sqrt{\delta_1} X_{1G},  \ \delta_1 \in[0,1],
\\ &X_2 = \sqrt{1-\delta_2} X_{2Gc} + \sqrt{\delta_2} X_{2Gp}, \ \delta_2 \in[0,1].
\\ &U_2 = X_{2Gc}, \ Q=\emptyset.
\end{align}
\label{eq:inputs scheme2}
\end{subequations}
In Scheme~II, $X_{2Gc}$ encodes a ``common'' message, and $X_{2Gp}$ and $X_{1G}$ encode the  ``private'' messages as in the classical Han-Kobayashi scheme \cite{H+K}. We shall also interpret 
$X_{1D}$ as encoding a ``common'' message even if $X_{1D}$ cannot be decoded at receiver~2 (the oblivious receiver) as receiver~2 lacks knowledge of the codebook(s) used by transmitter 1. 
The main message of the paper is in fact that, even with lack of codebook knowledge, if there would-be-common message is from a discrete alphabet then its effect on the rate region---up to a constant gap---is as if the message could indeed be jointly decoded.  We believe this is  because lack of codebook knowledge may be translated as lack of knowledge of the mapping of the codewords to the messages, but does not preclude a receiver's ability to perform {\it symbol-by-symbol  estimation} of the symbols in the interfering codeword (rather than decoding the messages carried by the codeword). Correctly estimating and subtracting off the interfering symbols is as effective 
as decoding the actual interfering 
codeword, as the message carried by the codeword is not desired anyhow. A similar intuition was pointed out in \cite{sand_decentr_proces} where the authors write ``We indeed see that BPSK signaling outperforms Gaussian signaling. This is because demodulation is some form of primitive decoding, which is not possible for the Gaussian signaling.''

In the next sections we will show that Proposition~\ref{prop:innerbound} with the inputs in~\eqref{eq:inputs scheme2} is gDoF optimal and is to within a constant gap of a capacity outer bound for the classical G-IC in the weak and moderate interference regimes. Also note that with $\delta_1=\delta_2=0$ Scheme~II in~\eqref{eq:inputs scheme2} reduces to Scheme~I in~\eqref{eq:inputs scheme1}.

The achievable region is derived for a general G-IC-OR and later on specialized to the symmetric case.
The rate region achievable by Scheme~II is:
\begin{thm}
\label{thm:sec:achScheme G-IC-OR:Ach2}
For the  G-IC-OR the following rate region is achievable with inputs as in \eqref{eq:inputs scheme2}
\begin{subequations}
\begin{align}
R_1 &\le 
 \mud\left(N, \frac{|h_{11}|^2(1-\delta_1)}{1+|h_{11}|^2\delta_1 + |h_{12}|^2\delta_2}  \right)
+\mug\left(\frac{|h_{11}|^2\delta_1}{1 + |h_{12}|^2\delta_2}\right),
\label{eq1:ach2:withScheme2}
\\
R_2 & \le
    \mud\left(N, \frac{|h_{21}|^2(1-\delta_1)}{1+|h_{21}|^2\delta_1+|h_{22}|^2} \right)
   +\mug\left(\frac{|h_{22}|^2}{1+|h_{21}|^2\delta_1}\right) \notag
\\&-\mug\left(\min\left(N^2-1,\frac{|h_{21}|^2(1-\delta_1)}{1+|h_{21}|^2\delta_1}\right)  \right),
\label{eq2:ach2:withScheme2}
\\
R_1+R_2 & \le 
    \mud\left(N, \frac{|h_{11}|^2(1-\delta_1)}{1+|h_{11}|^2\delta_1 + |h_{12}|^2}  \right)
   +\mug\left(|h_{11}|^2\delta_1 + |h_{12}|^2 \right)
   -\mug\left(|h_{12}|^2\delta_2 \right) \notag
\\&+\mud\left(N, \frac{|h_{21}|^2(1-\delta_1)}{1+|h_{21}|^2\delta_1 + |h_{22}|^2\delta_2}  \right)
   +\mug\left(\frac{ |h_{22}|^2\delta_2}{1+|h_{21}|^2\delta_1} \right) \notag
\\&-\mug\left( \min\left(N^2-1, \frac{|h_{21}|^2(1-\delta_1)}{1+|h_{21}|^2\delta_1}\right)  \right).
\label{eq3:ach2:withScheme2}
\end{align}
\label{eqall:ach2:withScheme2}
\end{subequations}
\end{thm}
\begin{IEEEproof}[Proof of Theorem~\ref{thm:sec:achScheme G-IC-OR:Ach2}]
The proof can be found in Appendix~\ref{app:proof thm:sec:achScheme G-IC-OR:Ach2} and follows similarly to the proof of  Theorem~\ref{thm:sec:achScheme G-IC-OR}.
\end{IEEEproof}

\section{High SNR performance}
\label{sec:gDoFach}
We now analyze the performance of the schemes in Theorems~\ref{thm:sec:achScheme G-IC-OR} and~\ref{thm:sec:achScheme G-IC-OR:Ach2} for the symmetric G-IC-OR at high-SNR by using the gDoF region as performance metric. 
The gDoF region is formally defined as follows. For an achievable pair $(R_1,R_2)$, let  
\begin{align}
\mathcal{D}(\alpha) := \left \{ (d_1,d_2) \in\mathbb{R}^2_+: 
d_i :=\lim_{\tiny
\begin{array}{l}
\inr=\snr^\alpha, \\
\snr \to \infty \\
\end{array}
} \frac{R_i}{\frac{1}{2}\log(1+\snr)},  i \in [1:2], \ (R_1,R_2) \ \text{is achievable}  \right\}.
\label{eq:low:gDofregion}
\end{align}

Let $\mathcal{D}^{\text{G-IC}}(\alpha)$ and $\mathcal{D}^{\text{G-IC-OR}}(\alpha)$ be the gDoF region of the classical G-IC and of the G-IC-OR, respectively.

We first present two different achievable gDoF regions based on Theorems~\ref{thm:sec:achScheme G-IC-OR} and~\ref{thm:sec:achScheme G-IC-OR:Ach2}, which we will compare to
$\mathcal{D}^{\text{G-IC}}(\alpha)$ given by \cite{etkin_tse_wang}
\begin{subequations}
\begin{align}
\mathcal{D}^{\text{G-IC}}(\alpha) \ : \
d_1 &\le 1,
\label{eq:gDoF sym GIC d1}\\
d_2 &\le 1,
\label{eq:gDoF sym GIC d2}\\
d_1+d_2 &\le \max(\alpha,2-\alpha),  
\label{eq:gDoF sym GIC d1d2 kra}\\ 
d_1+d_2 &\le \max(2\alpha,2-2\alpha),
\label{eq:gDoF sym GIC d1d2 etw}\\ 
2d_1+d_2&\le 2, \ \text{only for $\alpha\in[1/2,1]$}, 
\label{eq:gDoF sym GIC 2d1d2}\\
d_1+2d_2&\le 2, \ \text{only for $\alpha\in[1/2,1]$}.
\label{eq:gDoF sym GIC d12d2}  
\end{align}
\label{eq:gDoF sym GIC}
\end{subequations}

\begin{cor}[gDoF region from achievable Scheme I]
\label{thm:ach:gDoF:Region:strongInteference}
Let $N=\points(\snr^{\beta}) $ for some $\beta \ge 0$ then
\begin{subequations}
\begin{align}
\mathcal{D}^{\rm I}(\alpha,\beta) \ : \
d_1 &\le  \min(\beta,1),
\\
d_2 &\le  \min(\beta,[\alpha-1]^+)+ 1-\min(\beta,\alpha), 
\label{eq:thm:ach:gDoF:Region d2}
\\
d_1+d_2 &\le  \min(\beta,[1-\alpha]^+) +\alpha.
\end{align}
\label{eq:ach.reg. Gaussian ICOR U2=X2G gdof}
\end{subequations}
for any $\beta \ge 0$.  By Theorem~\ref{thm:sec:achScheme G-IC-OR}, the gDoF region $\mathcal{D}^{\rm I}(\alpha,\beta)$ is achievable.
\end{cor}
\begin{IEEEproof}[Proof of Corollary~\ref{thm:ach:gDoF:Region:strongInteference}]
We prove the bound 
in~\eqref{eq:thm:ach:gDoF:Region d2} only as the other bounds follow similarly. 
With $\inr=\snr^\alpha$ and $N=\points(\snr^{\beta})$ we have
\begin{align*}
&\lim_{\snr \to \infty}\frac{\log(N^2)}{\log(1+\snr)}    = \beta, \quad
\\
&\lim_{\snr \to \infty}\frac{\log(1+\inr)}{\log(1+\snr)} = \alpha.
\end{align*}
Therefore $d_2$ can be bounded as
\begin{align*}
d_2
  &=\lim_{\snr \to \infty} \frac{\text{left hand side of eq.\eqref{eq:ach.reg. Gaussian ICOR U2=X2 R2}}}{\frac{1}{2}\log(1+\snr)}
\\&=\min(\beta,[\alpha-1]^+)+1 - \min(\beta,\alpha),
\end{align*}
thus proving~\eqref{eq:thm:ach:gDoF:Region d2}.
\end{IEEEproof}

Next, by using Theorem~\ref{thm:sec:achScheme G-IC-OR:Ach2} with the power split as in \cite{etkin_tse_wang} we show yet another achievable gDoF region.
\begin{cor}[gDoF region from achievable Scheme II]
\label{thm:ach:gDoF:Region:weakInteference}
Let $N=\points(\snr^{\beta}) $ for some $\beta \ge 0$.
\begin{subequations}
\begin{align}
\mathcal{D}^{\rm II}(\alpha,\beta) \ : \
d_1 &\le  \min(\beta,1+\alpha-\max(1,\alpha))+[1-\alpha]^+,
\\
d_2 &\le  \min(\beta,[\alpha-1]^+)+1 - \min(\beta,\alpha),
\\
d_1+d_2 &\le  \min(\beta,[1+\alpha-\max(1,2\alpha)]^+)+ \max(\alpha,1-\alpha)+ \notag
\\&+          \min(\beta,[2\alpha-\max(1,\alpha)]^+)+[1-\alpha]^+- \min(\beta,\alpha).
\end{align}
\end{subequations}
for any $\beta\geq 0$.  By Theorem~\ref{thm:sec:achScheme G-IC-OR:Ach2}, the gDoF region $\mathcal{D}^{\rm II}(\alpha,\beta)$ is achievable.

\end{cor}
\begin{IEEEproof}[Proof of Corollary~\ref{thm:ach:gDoF:Region:weakInteference}]
Let $\inr=\snr^\alpha$, $N=\points(\snr^{\beta})$, and $\delta_1=\delta_2=\frac{1}{1+\inr}$ in Theorem~\ref{thm:sec:achScheme G-IC-OR:Ach2} (see the region in~\eqref{eqall:ach2:etw sym} in Appendix~\ref{app:proof thm:sec:achScheme G-IC-OR:Ach2}) and take limits similarly to the proof of Corollary~\ref{thm:ach:gDoF:Region:strongInteference}.
\end{IEEEproof}

We are now ready to prove the main result of this section:
\begin{thm}
\label{thm:GDOFregion}
For the G-IC-OR there is no loss in gDoF compared to the classical G-IC, i.e.,
\[
\mathcal{D}^{\text{G-IC}}(\alpha)=\mathcal{D}^{\text{G-IC-OR}}(\alpha).
\]
\end{thm}

\begin{figure*}
\centering
\subfigure[Very strong interference.] {
\includegraphics[width=6.5cm]{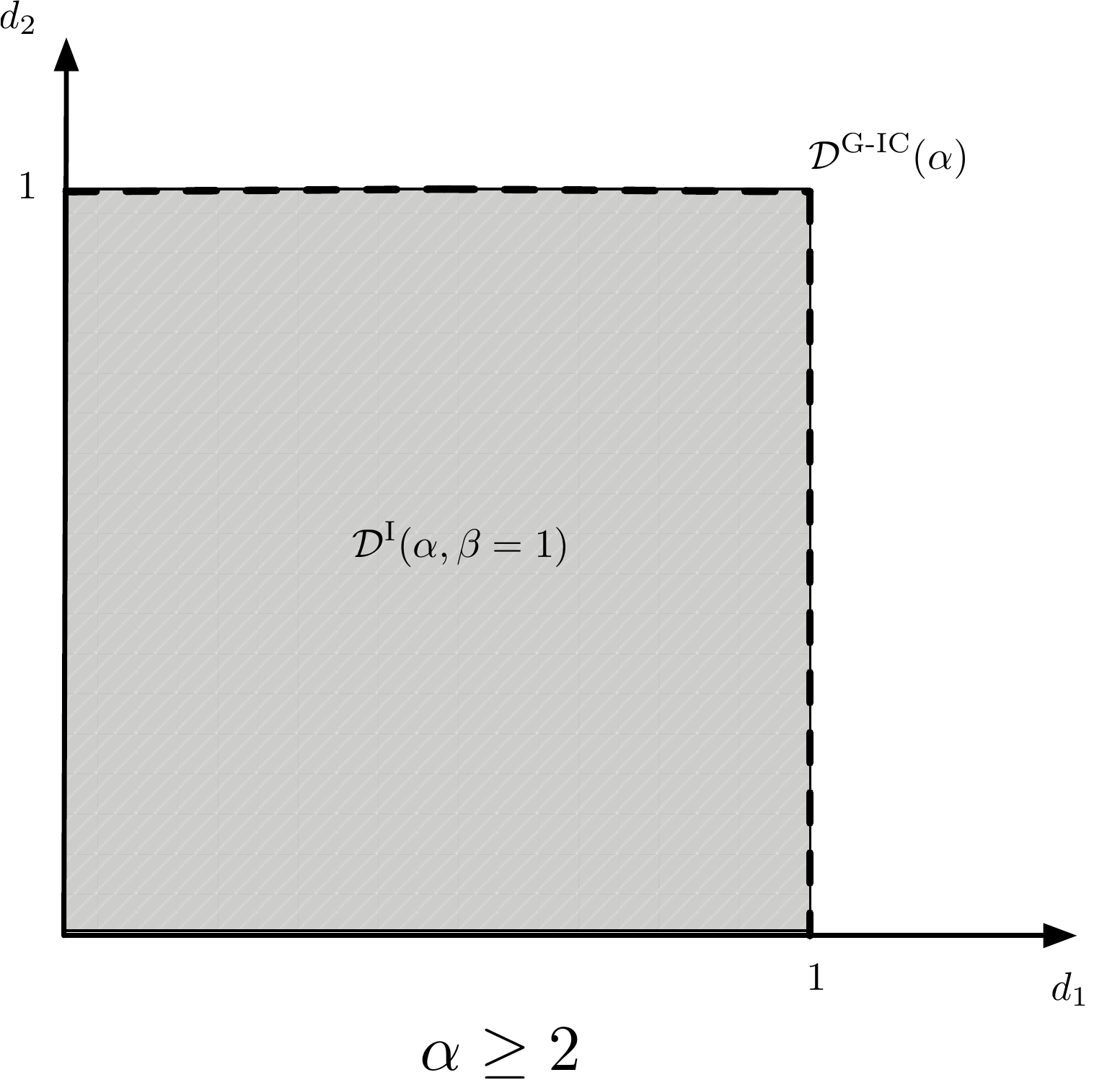}
\label{fig:gDoFachVERYStrong}
}
\hfill
\subfigure[Strong interference.] {
\includegraphics[width=7cm]{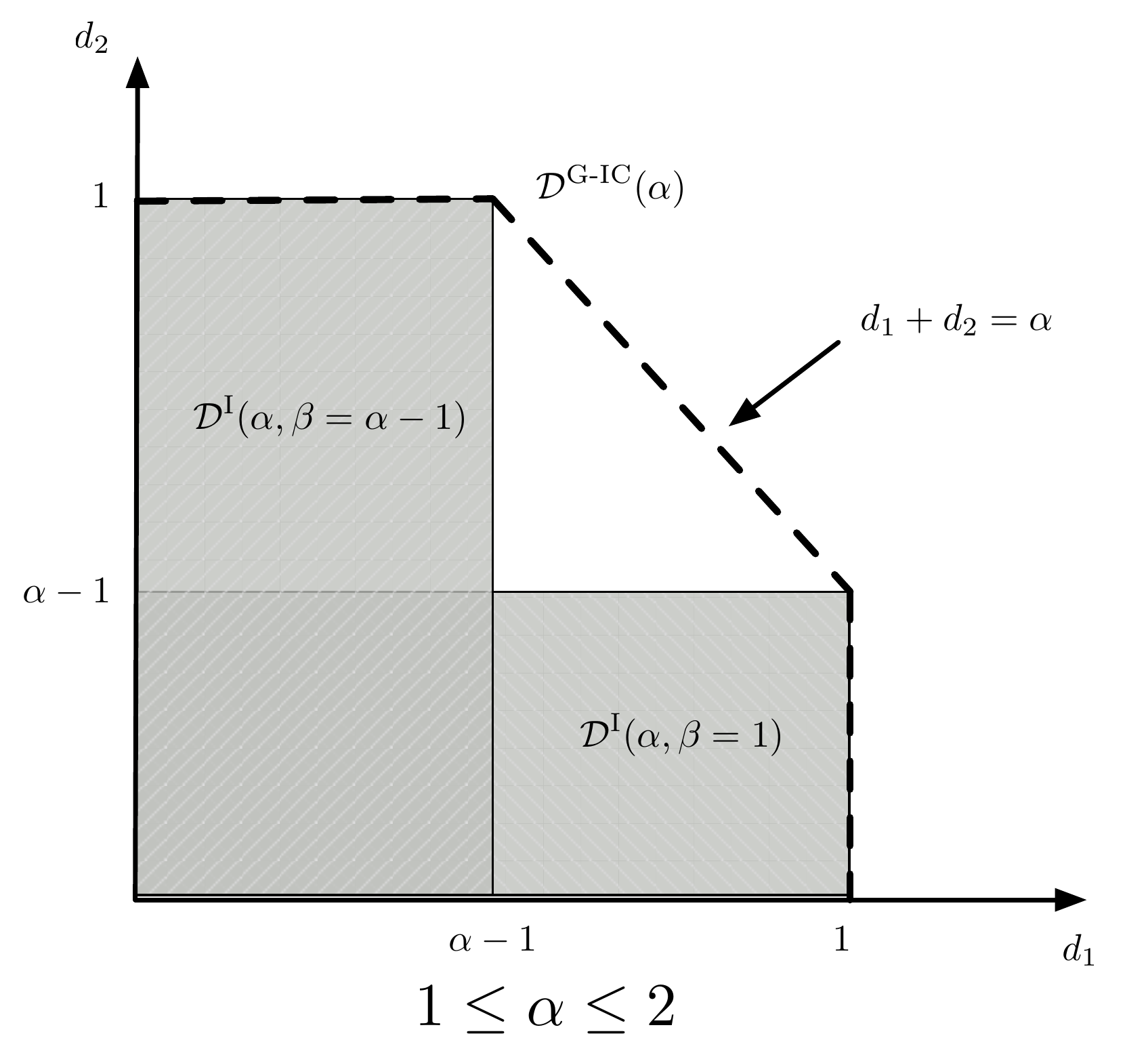}
\label{fig:gDoFachStrong}
}
\hfill
\subfigure[Moderately weak interference~1.] {
\includegraphics[width=7cm]{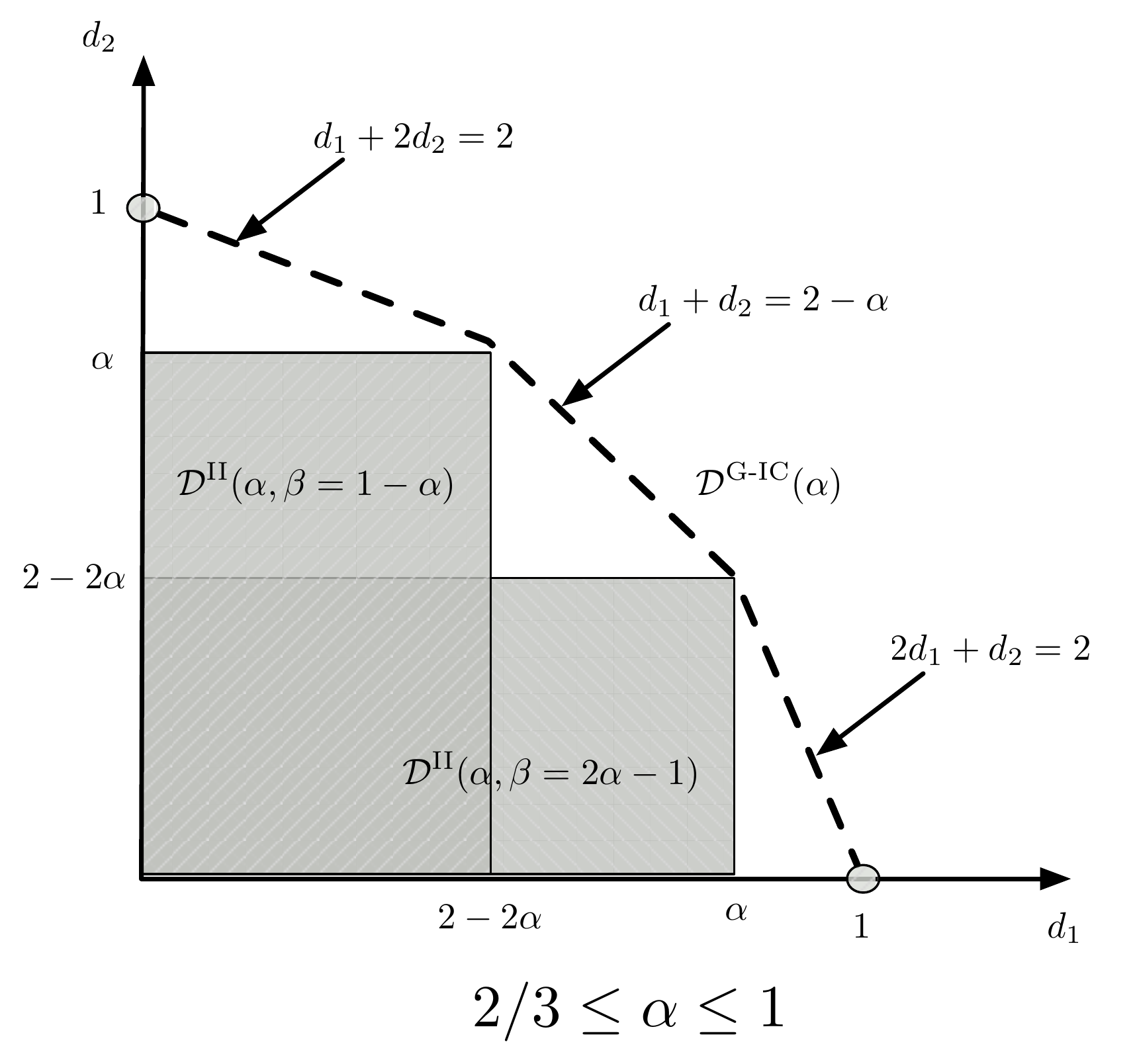}
\label{fig:gDoFachModerate}
}
\hfill
\subfigure[Moderately weak interference~2.] {
\includegraphics[width=7cm]{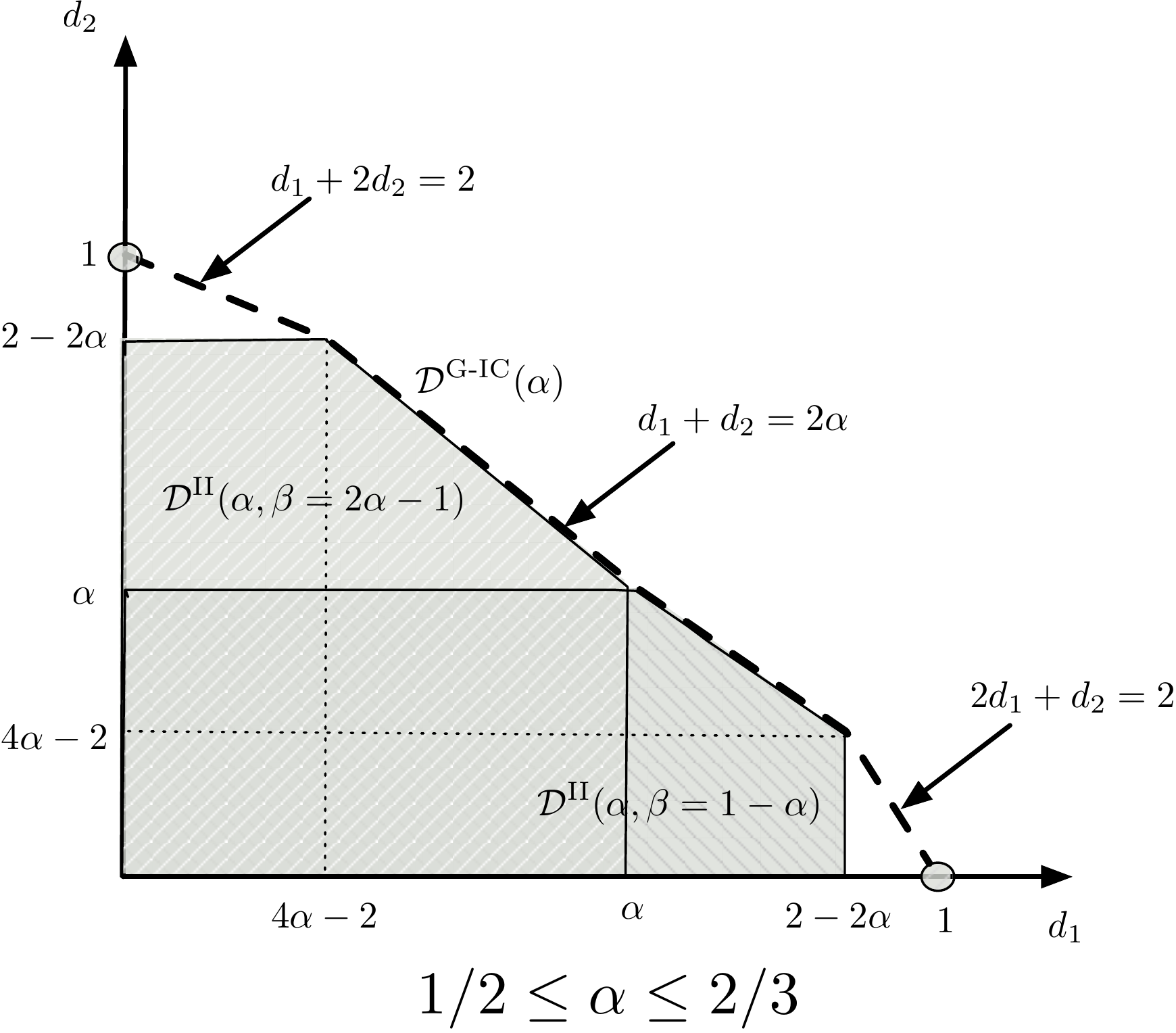}
\label{fig:gDoFachWeak}
}
\caption{How to achieve the gDoF region for the G-IC-OR in different parameter regimes. }
\end{figure*}

\begin{IEEEproof}[Proof of Theorem~\ref{thm:GDOFregion}]
We consider several regimes:

\paragraph{Very strong interference regime $\alpha\geq 2$}
In this regime the gDoG region outer bound $\mathcal{D}^{\text{G-IC}}(\alpha)$ is characterized by~\eqref{eq:gDoF sym GIC d1} and~\eqref{eq:gDoF sym GIC d2}.
For achievability we consider Corollary~\ref{thm:ach:gDoF:Region:strongInteference} with $\beta=1$, that is,
\begin{align*}
\mathcal{D}^{\rm I}(\alpha,1) \ : \
d_1 &\le  \min(1,1) = 1,
\\
d_2 &\le  \min(1,[\alpha-1]^+)+1 - \min(1,\alpha) = 1,
\\\
d_1+d_2 &\le  \min(1,[1-\alpha]^+) +\alpha = \alpha  \ (\text{redundant because $\alpha\geq 2$}).
\end{align*}
Since the sum-gDoF is redundant, we get that 
\[
\mathcal{D}^{\rm I}(\alpha,\beta=1) = \{d_i\in[0,1], \ i\in[1:2]\}
=\mathcal{D}^{\text{G-IC-OR}}(\alpha)
=\mathcal{D}^{\text{G-IC}}(\alpha).
\]
Fig.~\ref{fig:gDoFachVERYStrong} illustrates the region $\mathcal{D}^{\rm I}(\alpha,\beta=1)$.

\paragraph{Strong interference regime $1 \leq \alpha < 2$}
In this regime the gDoG region outer bound $\mathcal{D}^{\text{G-IC}}(\alpha)$ is characterized by~\eqref{eq:gDoF sym GIC d1}-\eqref{eq:gDoF sym GIC d1d2 kra} and has two dominant corner points: $(d_1,d_2)=(1,\alpha-1)$ and $(d_1,d_2)=(\alpha-1,1)$.
For achievability we consider the following achievable gDoF regions
\begin{align*}
\mathcal{D}^{\rm I}(\alpha,1) \ : \  
d_1 &\le 1,\\
d_2 &\le \alpha-1, \\
d_1+d_2 &\le \alpha \ (\text{redundant}).
\end{align*}
and
\begin{align*}
\mathcal{D}^{\rm I}(\alpha,\alpha-1) \ : \
d_1 & \le \alpha-1,\\
d_2 & \le 1,\\
d_1+d_2 & \le \alpha \ (\text{redundant}),
\end{align*}
Fig.~\ref{fig:gDoFachStrong} illustrates that
\[
\co \left( \mathcal{D}^{\rm I}(\alpha,1)  \cup  \mathcal{D}^{\rm I}(\alpha,\alpha-1) \right)
=
\mathcal{D}^{\text{G-IC}}(\alpha)
=
\mathcal{D}^{\text{G-IC-OR}}(\alpha).
\]

\paragraph{Moderately weak interference regime $\frac{1}{2} < \alpha  < 1$} 
In this regime the gDoG region outer bound $\mathcal{D}^{\text{G-IC}}(\alpha)$ is characterized by all the constraints in~\eqref{eq:gDoF sym GIC} and has four corner points: $(d_1,d_2)=(1,0)$, $(d_1,d_2)=(0,1)$, and 
$(d_1,d_2)=(\min(4\alpha-2,\alpha),2-2\alpha)$ and $(d_1,d_2)=(2-2\alpha,\min(4\alpha-2,\alpha))$. 
The gDoF pair $(d_1,d_2)=(1,0)$ is trivially achievable by silencing user 2,  
and similarly $(d_1,d_2)=(0,1)$ by silencing user 1. 
For achievability of the remaining two corner points, 
we consider the following achievable gDoF regions
\begin{align*}
\mathcal{D}^{\rm II}(\alpha,2\alpha-1) \ : \
d_1 &\le  \min(2\alpha-1,1+\alpha-1)+1-\alpha 
= \alpha,
\\
d_2 &\le  \min(2\alpha-1,0)+1 - \min(2\alpha-1,\alpha) 
=2-2\alpha,
\\
d_1+d_2 &\le  \min(2\alpha-1,[1+\alpha-\max(1,2\alpha)]^+)+ \max(\alpha,1-\alpha)+ \notag
\\&+          \min(2\alpha-1,[2\alpha-1]^+)+1-\alpha- \min(2\alpha-1,\alpha) \notag
\\&=\min(2\alpha,2-\alpha), \quad \text{(redundant for $\alpha\in[2/3,1]$)}.
\end{align*}
and
\begin{align*}
\mathcal{D}^{\rm II}(\alpha,1-\alpha) \ : \
d_1 &\le  \min(1-\alpha,1+\alpha-1)+1-\alpha 
=2-2\alpha,
\\
d_2 &\le  \min(1-\alpha,0)+1 - \min(1-\alpha,\alpha) 
= \alpha,
\\
d_1+d_2 &\le  \min(1-\alpha,[1+\alpha-\max(1,2\alpha)]^+)+ \max(\alpha,1-\alpha)+ \notag
\\&+          \min(1-\alpha,[2\alpha-1]^+)+1-\alpha- \min(1-\alpha,\alpha) \notag
\\&=\min(2\alpha,2-\alpha),  \quad \text{(redundant for $\alpha\in[2/3,1]$)}.
\end{align*}

Fig.~\ref{fig:gDoFachModerate} (for $\alpha\in[2/3,1]$) and
Fig.~\ref{fig:gDoFachWeak} (for $\alpha\in[1/2,2/3]$) illustrate that
\begin{align*}
&\co \left( \big\{(d_1,d_2)=(1,0)\big\} \cup \big\{(d_1,d_2)=(0,1)\big\} \cup
\mathcal{D}^{\rm II}(\alpha,2\alpha-1) \cup \mathcal{D}^{\rm II}(\alpha,1-\alpha) \right)
=
\mathcal{D}^{\text{G-IC}}(\alpha) \\
&=
\mathcal{D}^{\text{G-IC-OR}}(\alpha).
\end{align*}

\paragraph{Noisy Interference $0\leq \alpha \le \frac{1}{2}$}  
In this regime one may achieve the whole optimal G-IC gDoF region by using Gaussian inputs, treating interference as noise, and power control. Since this strategy is feasible for the G-IC-OR, the G-IC gDoF region is achievable also for the G-IC-OR.

This concludes our proof. 
\end{IEEEproof}

The result of Theorem~\ref{thm:GDOFregion} is quite surprising, namely, that for the G-IC-OR we can achieve the gDoF region of the classical G-IC in all regimes. This is especially surprising in the strong and very strong interference regimes where joint decoding of intended and interfering messages is optimal for the classical G-IC---recall that joint decoding appears to be precluded by the absence of codebook knowledge in the G-IC-OR. 
This seems to suggest that while decoding of the undesired messages is not possible, one may still estimate (i.e., symbol-by-symbol demodulate) the codeword symbols corresponding to the undesired messages.

\section{Finite SNR performance}

\label{sec:finiteSNR}
In the previous section we showed that the gDoF region of the classical G-IC can be achieved even when one receiver lacks knowledge of the interfering codebook. One may then ask whether it is possible to achieve the capacity, possibly up to a constant gap, of the classical G-IC at all finite SNRs.  We next show that this is indeed possible. 
For future use, the capacity region of the classical G-IC is outer bounded by~\cite{etkin_tse_wang}
\begin{subequations}
\begin{align}
\mathcal{R}_{\rm out}^{\text{(G-IC)}} : \quad
R_1 &\leq \mug\left(\snr\right), 
\label{eq:etw outer bound:r1}
\\
R_2 &\leq \mug\left(\snr\right), 
\label{eq:etw outer bound:r2}
\\
R_1+R_2 &\leq \Big[\mug\left(\snr\right)-\mug\left(\inr\right)\Big]^+ +\mug\left(\snr+\inr\right),
\label{eq:etw outer bound:r1+r2 kra}
\\
R_1+R_2 &\leq 2\mug\left(\inr+\frac{\snr}{1+\inr}\right),
\label{eq:etw outer bound:r1+r2 etw}
\\
2R_1+R_2 &\leq \Big[\mug\left(\snr\right)-\mug\left(\inr\right)\Big]^+ 
+\mug\left(\snr+\inr\right)+\mug\left(\inr+\frac{\snr}{1+\inr}\right), 
\label{eq:etw outer bound:2r1+r2}
\\
R_1+2R_2 &\leq \Big[\mug\left(\snr\right)-\mug\left(\inr\right)\Big]^+
+\mug\left(\snr+\inr\right)+\mug\left(\inr+\frac{\snr}{1+\inr}\right),
\label{eq:etw outer bound:r1+2r2}
\end{align}
\label{eq:etw outer bound:all}
\end{subequations}
which is tight for $\snr\leq \inr$ and optimal to with $1/2$~bit (per channel use per user) otherwise.

The main result of this section is:
\begin{thm}
\label{thm:cosntant gap}
For the G-IC-OR it is possible to achieve the outer bound region in~\eqref{eq:etw outer bound:all} to within \gaptoclassicalIC~bits per channel use per user.
\end{thm}
\begin{IEEEproof}[Proof of Theorem~\ref{thm:cosntant gap}]
We consider different regimes separately.

\setcounter{paragraph}{0}

\paragraph{Very strong interference $\snr(1+{\snr})\leq\inr$}
In the regime the capacity region of the classical G-IC is given by~\eqref{eq:etw outer bound:r1} and~\eqref{eq:etw outer bound:r2}.
For achievability we consider the achievable region in Theorem~\ref{thm:sec:achScheme G-IC-OR} with
\begin{align}
&N = \points(\snr) \quad \text{(equivalent of $\beta = 1$)}
\notag\\&
\Longrightarrow  N^2-1 \leq \snr \leq \frac{\inr}{1+\snr}\leq \inr.
\label{eq: paramfor gap very strong}
\end{align}
Recall that the achievable region in Theorem~\ref{thm:sec:achScheme G-IC-OR} is the region in~\eqref{eq:inner for one Rx oblivious} with the inputs as in~\eqref{eq:inputs scheme1}; the sum-rate in Theorem~\ref{thm:sec:achScheme G-IC-OR} is redundant if $I(X_1;Y_1|X_2)+I(X_2;Y_2) \leq I(X_1,X_2;Y_1)$, that is, if 
$I(X_2;Y_2) \leq I(X_2;Y_1)$, for all input distributions in~\eqref{eq:inputs scheme1}.
With a Gaussian $X_2$ as in~\eqref{eq:inputs scheme1}: 
\[
I(X_2;Y_2) \leq I(X_2;Y_2|X_1) 
= I(X_{2G}; \sqrt{\snr} \ X_{2G}+Z_2)= \mug(\snr),
\]
and
\[
I(X_2;Y_1) = I(X_{2G}; \sqrt{\inr} \ X_{2G}+\sqrt{\snr} \ X_{1D}+Z_2) \geq \mug\left(\frac{\inr}{1+\snr}\right),
\]
because a Gaussian noise is the worst noise for a Gaussian input. 
Since in very strong interference we have $\mug(\snr) \leq \mug\left(\frac{\inr}{1+\snr}\right)$,
the condition  $I(X_2;Y_2) \leq I(X_2;Y_1)$ is verified for all inputs in~\eqref{eq:inputs scheme1} and hence
we can drop the sum-rate constraint in~\eqref{eq:ach.reg. Gaussian ICOR U2=X2 R1+R2} 
from Theorem~\ref{thm:sec:achScheme G-IC-OR}. Therefore, in this regime
the following rates are achievable
\begin{subequations}
\begin{align}
\mathcal{R}_{\rm in}^{\text{(G-IC-OR very strong)}} : \quad
R_1 &\leq \mug(\snr) - \Delta_1, \label{eq:gap G-IC-OR very strong in R1 in}
\\
R_2 &\leq \mug(\snr) - \Delta_2, \label{eq:gap G-IC-OR very strong in R2 in}
\end{align}
where for $N = \points(\snr)$
\begin{align}
\Delta_1 
  &:= \mug(\snr)-\mud\left(N, \snr\right)\notag
\\&\leq \frac{1}{2}\log\left(\frac{4\pi\eu}{3}\right) \ \text{for the reasoning leading to~eq.\eqref{eq:shaping and integer losses}},  
\label{eq:gap G-IC-OR very strong in R1 gap} 
\\
\Delta_2 &:= \mug\left(\min \left(N^2-1, \inr \right)\right)-\mud\left(N,  \frac{\inr}{1+\snr}\right)\notag
\\&= \log(N)-\left[\log(N)-\frac{1}{2}\log\left(\frac{\pi\eu}{3}\right)\right]^+
\leq \frac{1}{2}\log\left(\frac{\pi\eu}{3}\right),
\label{eq:gap G-IC-OR very strong in R2 gap}
\end{align}
\end{subequations}
where the equality in~\eqref{eq:gap G-IC-OR very strong in R2 gap} is a consequence of the relationships in~\eqref{eq: paramfor gap very strong}.

It is immediate to see that~\eqref{eq:gap G-IC-OR very strong in R1 gap} is the gap for $R_1$ 
and that~\eqref{eq:gap G-IC-OR very strong in R2 gap} is the gap for $R_2$.
Therefore in this regime the gap is at most $\frac{1}{2}\log\left(\frac{4\pi\eu}{3}\right)$ per channel use per user, and it is due to shaping loss and integer penalty.

\paragraph{Strong interference $\snr\leq\inr < \snr(1+{\snr})$}
In this regime the capacity region of the classical G-IC is given by~\eqref{eq:etw outer bound:r1}-\eqref{eq:etw outer bound:r1+r2 kra}, and has two dominant corner points
\begin{subequations}
\begin{align}
\mathcal{R}_{\rm out}^{\text{(G-IC strong P1)}} : \quad
\left(R_1,R_2\right) = \left( \mug\left(\snr\right), \mug\left(\frac{\inr}{1+\snr}\right) \right),
\label{eq:CornerPoint1Strong}
\end{align}
and 
\begin{align}
\mathcal{R}_{\rm out}^{\text{(G-IC strong P2)}} : \quad
\left(R_1,R_2\right) = \left( \mug\left(\frac{\inr}{1+\snr}\right), \mug\left(\snr\right) \right).
\label{eq:CornerPoint2Strong}
\end{align}
\end{subequations}
The other two corner points are $(R_1,R_2)=(\mug(\snr),0)$ and $(R_1,R_2)=(0,\mug(\snr))$ that can be exactly achieved by silencing one of the users.

For achievability we mimic the proof of the gDoF region in the same regime (see Fig.~\ref{fig:gDoFachStrong}), that is, we show the achievability to within a constant gap of the corner points in~\eqref{eq:CornerPoint1Strong} and~\eqref{eq:CornerPoint2Strong} by choosing two different values of $N$ in Theorem~\ref{thm:sec:achScheme G-IC-OR}.
For the corner point in \eqref{eq:CornerPoint1Strong} we consider the achievable region in Theorem~\ref{thm:sec:achScheme G-IC-OR} with
\begin{subequations}
\begin{align}
&N =\points(\snr) \quad \text{(equivalent of $\beta = 1$)}
\notag\\&
\Longrightarrow N^2-1\leq \snr \leq \inr \leq \snr(1+\snr),
\label{eq:NpointCornerPoint1Strong}
\end{align}
and for the corner point \eqref{eq:CornerPoint2Strong} we consider the achievable region in Theorem~\ref{thm:sec:achScheme G-IC-OR} with
\begin{align}
&N =\points\left(\frac{\inr}{1+\snr}\right) \quad \text{(equivalent of $\beta = \alpha-1$)}
\notag\\&
\Longrightarrow N^2-1\leq \frac{\inr}{1+\snr} \leq \snr \leq \inr.
\label{eq:NpointCornerPoint2Strong}
\end{align}
\end{subequations}

For the choice of $N$ in~\eqref{eq:NpointCornerPoint1Strong} the achievable region in Theorem~\ref{thm:sec:achScheme G-IC-OR} can be written as 
\begin{align*}
R_1 
  &\leq \mud\left(N, \snr \right) 
\\&= \left[\log\left(N \right)-\frac{1}{2}\log\left(\frac{\pi\eu}{3}\right)\right]^+,
\\
R_2 
  &\leq\mud\left(N, \frac{\inr}{1+\snr}\right) + \mug\left(\snr\right)
         -\mug\left( \min \left(N^2-1,\inr \right) \right)
\\&=\left[\mug\left(\frac{\inr}{1+\snr} \right)-\frac{1}{2}\log\left(\frac{\pi\eu}{3}\right)\right]^+
+\mug(\snr)- \log(N),
\\
R_1+R_2 &\leq\mud\left(N, \frac{\snr}{1+\inr}\right) + \mug\left(\inr\right)
\\&=\left[\mug\left(\frac{\snr}{1+\inr} \right)-\frac{1}{2}\log\left(\frac{\pi\eu}{3}\right)\right]^++\mug(\inr),
\end{align*}
which can further be lower bounded as 
\begin{subequations}
\begin{align}
\mathcal{R}_{\rm in}^{\text{(G-IC strong P1)}} \  : \
R_1 &\leq \log\left(N \right)-\frac{1}{2}\log\left(\frac{\pi\eu}{3}\right) \notag
\\&= \mug\left(\snr\right) - \Delta_1,
\\
R_2 &\leq \mug\left(\snr+\inr\right) -\log(N) -\frac{1}{2}\log\left(\frac{\pi\eu}{3}\right)\notag
\\&= \mug\left(\frac{\inr}{1+\snr}\right) - \Delta_2,
\\
R_1+R_2 &\leq \mug\left(\snr+\inr\right)-\frac{1}{2}\log\left(\frac{\pi\eu}{3}\right) \notag
\\&= \left(\mug\left(\snr\right) - \Delta_1\right)+\left(\mug\left(\frac{\inr}{1+\snr}\right) - \Delta_2\right)+\frac{1}{2}\log\left(\frac{\pi\eu}{3}\right),
\end{align}
\end{subequations}
where the sum-rate bound is clearly redundant and where 
\begin{subequations}
\begin{align}
\Delta_1 &:=\mug\left(\snr\right)-\log(N) +\frac{1}{2}\log\left(\frac{\pi\eu}{3}\right) 
\leq \frac{1}{2}\log\left(\frac{4\pi\eu}{3}\right),
\\
\Delta_2 &:=\log(N)-\mug\left(\snr\right) +\frac{1}{2}\log\left(\frac{\pi\eu}{3}\right)
\leq \frac{1}{2}\log\left(\frac{\pi\eu}{3}\right).
\end{align}
\label{eq:AchStrongSimp}
\end{subequations}
Therefore, with $N$ as in~\eqref{eq:NpointCornerPoint1Strong} in Theorem~\ref{thm:sec:achScheme G-IC-OR}, the gap to the corner point in~\eqref{eq:CornerPoint1Strong} is at most $\frac{1}{2}\log\left(\frac{4\pi\eu}{3}\right)$ per channel use per user, as for the very strong interference regime.

By following similar steps, for the choice of $N$ in~\eqref{eq:NpointCornerPoint2Strong} in Theorem~\ref{thm:sec:achScheme G-IC-OR},
the gap to the corner point in~\eqref{eq:CornerPoint2Strong} is still given by~\eqref{eq:AchStrongSimp}, that is, the gap is at most $\frac{1}{2}\log\left(\frac{4\pi\eu}{3}\right)$ per channel use per user, as for the very strong interference regime.

\paragraph{Moderately weak interference $\inr  \le \snr \le \inr(1+\inr)$}
In this regime the capacity of the G-IC is outer bounded by~\eqref{eq:etw outer bound:all}.

As we did for the gDoF region (see Figs.~\ref{fig:gDoFachModerate} and~\ref{fig:gDoFachWeak}), we show here that we can achieve, up to a constant gap, all dominant corner points of~\eqref{eq:etw outer bound:all}.
By silencing one of the users, we can achieve $(R_1,R_2)=(\mug(\snr),0)$ and $(R_1,R_2)=(0,\mug(\snr))$; these rate points are to within 1~bit of the corner points of~\eqref{eq:etw outer bound:all} given by $(R_1,R_2)=\left(A,\mug\left(\snr\right)\right)$  and $(R_1,R_2)=\left(\mug\left(\snr\right),A\right)$ where
\begin{align*}
A &:=
 \mug\left(\snr+\inr\right)
+\mug\left(\inr+\frac{\snr}{1+\inr}\right)
-\mug\left(\snr\right)
-\mug\left(\inr\right)
\\&=
 \mug\left(\frac{\inr}{1+\snr}\right)
+\mug\left(\frac{\snr}{(1+\inr)^2}\right)
\\&\leq 
 \mug\left(\frac{\snr}{1+\snr}\right)
+\mug\left(\frac{\inr}{1+\inr}\right)
\leq 2\cdot \frac{1}{2}\log(2) = 1.
\end{align*}

We therefore have to show the achievability of the remaining two corner points obtained by the intersection of the sum-rate outer bound (given by $\min$(eq.\eqref{eq:etw outer bound:r1+r2 kra},eq.\eqref{eq:etw outer bound:r1+r2 etw})) with either~\eqref{eq:etw outer bound:2r1+r2} or~\eqref{eq:etw outer bound:r1+2r2}. 
For these corner points, the gDoF-optimal choices of $\beta$ were $2\alpha-1$ and $1-\alpha$, which we mimic here by choosing  the following values of $N$ in the region in~\eqref{eqall:ach2:etw sym} (a simplified achievable region from Theorem~\ref{thm:sec:achScheme G-IC-OR:Ach2})
\begin{align}
&N =\points\left(\frac{\inr^2}{1+\snr+2\inr}\right) \quad \text{(equivalent of $\beta = 2\alpha-1$)}
\notag\\&
\Longrightarrow 
N^2-1 \leq \frac{\inr^2}{1+\snr+2\inr} 
      \leq \min\left(\frac{\inr^2}{1+2\inr},\frac{\inr \cdot \snr}{1+\snr+2\inr}\right),
\label{eq:NpointCornerPoint1weak}
\end{align}
because $\inr \leq \snr$,
and 
\begin{align}
&N =\points\left(\frac{\snr \cdot \inr }{(1+\inr)^2+\snr}\right) \quad \text{(equivalent of $\beta = 1-\alpha$)}
\notag\\&
\Longrightarrow 
N^2-1 \leq \frac{\snr \cdot \inr }{(1+\inr)^2+\snr}
      \leq \min\left(\frac{\inr^2}{1+2\inr},\frac{\inr \cdot \snr}{1+\snr+2\inr}\right),
\label{eq:NpointCornerPoint2weak}
\end{align}
because $\snr \le \inr(1+\inr)$.
In the regime $\inr \leq \snr \le \inr(1+\inr)$ we also have
\begin{align}
\frac{\inr^2}{(1+\inr)(1+\snr)+\inr} \leq \frac{\inr^2}{(1+\inr)^2+\inr} \leq 1 \leq N^2-1, \quad \forall N\geq 2.
\label{eq:someothercoditioninweak}
\end{align}
With~\eqref{eq:NpointCornerPoint1weak}-\eqref{eq:someothercoditioninweak}, and by recalling that
$\mug(x)-\frac{1}{2}\log(4) \leq \log(\points(x)) \leq \mug(x), \ x\geq 0$,
the region in~\eqref{eqall:ach2:etw sym} can be further lower bounded as follows%
\footnote{
In order to get the sum-rate, let $n=N^2-1\in\mathbb{N}$ and consider either 
$N=\points(a) : n_a:=\points(a)^2-1\leq a \in\mathbb{R}^+$ or 
$N=\points(b) : n_b:=\points(a)^2-1\leq b\in\mathbb{R}^+$ in the expression 
$y(n) := \mug(\min(n,a))+\mug(\min(n,b))-\mug(n)$ that appears in the sum-rate.
It follows easily that for 
$N=\points(a) \ : \ y= \mug(\min(n_a,b)) \geq \mug(\min(n_a,n_b)) \geq \mug(\min(a,b))-\frac{1}{2}\log(4)$,
and for 
$N=\points(b) \ : \ y= \mug(\min(a,n_b)) \geq \mug(\min(n_a,n_b)) \geq \mug(\min(a,b))-\frac{1}{2}\log(4)$,
where the term $\frac{1}{2}\log(4)$ is due to the ``integer penalty''.
}
\begin{subequations}
\begin{align}
\mathcal{R}_{\rm in}^{\text{(G-IC-OR weak)}} : \quad
R_1 &\le
    {\mug(x) -\frac{1}{2}\log(4)}
\notag\\& - \frac{1}{2}\log\left(\frac{\pi\eu}{3}\right) +\mug\left(\frac{\snr}{1+2\inr}\right),
\label{eq1:ach2:etw sym weak}\\
R_2 & \le
   {\mug\left(\frac{\inr^2}{(1+\inr)(1+\snr)+\inr}\right)- \frac{1}{2}\log\left(\frac{\pi\eu}{3}\right)}     
\notag\\&
   +\mug\left(\frac{\snr}{2}\right)
   {-\mug(x),}
\label{eq2:ach2:etw sym weak}\\
R_1+R_2 & \le
    {\mug\left(\min\left( \frac{\inr^2}{1+\snr+2\inr}, \frac{\snr \cdot \inr }{(1+\inr)^2+\snr} \right)\right)  -\frac{1}{2}\log(4)}
\notag
\\&+\mug\left(\inr+\frac{\snr}{1+\inr}\right)
   -\mug\left(\frac{\inr}{1+\inr}\right)
   +\mug\left(\frac{\snr}{1+2\inr}\right)
\notag
\\&-2\cdot \frac{1}{2}\log\left(\frac{\pi\eu}{3}\right),    
\label{eq3:ach2:etw sym weak}
\end{align}
where
\begin{align}
x &:= \frac{\inr^2}{1+\snr+2\inr}              \ \text{if $N$ as in~\eqref{eq:NpointCornerPoint1weak}, or } 
\label{eq:x for NpointCornerPoint1weak}
\\
x &:= \frac{\snr \cdot \inr }{(1+\inr)^2+\snr} \ \text{if $N$ as in~\eqref{eq:NpointCornerPoint2weak}}. 
\label{eq:x for NpointCornerPoint2weak}
\end{align}
\label{eqall:ach2:etw sym weak}
\end{subequations}
In Appendix~\ref{app:gap moderately week tedious} we show that region in~\eqref{eqall:ach2:etw sym weak} achieves the classical G-IC outer bound to within \gaptoclassicalIC~bits (per channel user per user).

\paragraph{Noisy interference $\inr(1+\inr)\leq \snr$}
In this regime Gaussian inputs, treating interference as noise, and power control is optimal to within $1/2$~bit (per channel use per user) for the classical G-IC; since this scheme does not require codebook knowledge / joint decoding, the gap result applies to the G-IC-OR as well.

This concludes the proof.
\end{IEEEproof}

\section{Conclusion} 
\label{sec:Conclusion}
In this paper we derived  capacity results for the interference channel where one of the receivers lacks knowledge of the interfering codebook, in contrast to a classical model where both receivers possess full codebook knowledge.
For the class of injective semi-deterministic interference channels with one oblivious receiver, we derived a capacity result to within a constant gap; the gap is zero for fully deterministic channels, thereby providing an exact capacity characterization.
We also derived the exact capacity region for a general memoryless interference channel with one oblivious receiver in the regime where the non-oblivious receiver experiences very strong interference.

We next proceeded to the Gaussian noise channel, where, 
unlike past work on oblivious receivers, we were able to demonstrate performance guarantees. For the symmetric case we derived the gDoF region and the capacity region to within a constant gap of \gaptoclassicalIC~bits (per channel use per user). Surprisingly, this lack of codebook knowledge at one receiver does not impact the gDoF at all, and only the Gaussian capacity region to within a constant gap, compared to having full codebook knowledge. 
We believe this is because even though the mapping from codewords to messages may not be known, this does not prevent the receiver from  estimating (for example by  symbol-by-symbol demodulation) and removing the effect of the interfering codeword itself.

{An interesting future direction is to consider a generalization  with lack of interfering codebook knowledge at both receivers, where one might surmise that both inputs would have discrete components. 
However, this generalization is highly non-trivial and significantly more mathematically challenging, and was left as an open problem in \cite{simion_codebook}.  
The major issue that arises when both users employ discrete inputs is the need to compute the cardinality and minimum distance of the sum of two discrete sets.  These quantities are not only difficult to compute in general, but are also very sensitive  to whether channel gains are rational or irrational (this is an open problem in additive combinatorics). For progress on this problem see our conference work \cite{DytsoITA2014,dytsoTINconf}, for which the journal version is currently under preparation. }

\appendices

\section{Proof of Theorem~\ref{thm:sec:achScheme G-IC-OR:Ach2}}
\label{app:proof thm:sec:achScheme G-IC-OR:Ach2}
We proceed to evaluate the rate region in Proposition~\ref{prop:innerbound} with the inputs in~\eqref{eq:inputs scheme2}. 
With the chosen inputs, the outputs are
\begin{align*}
&Y_1=h_{11}\sqrt{1-\delta_1} X_{1D}  + h_{11}\sqrt{\delta_1} X_{1G}+h_{12}\sqrt{1-\delta_2} X_{2Gc} + h_{12}\sqrt{\delta_2} X_{2Gp}+Z_1,
\\
&Y_2=h_{21}\sqrt{1-\delta_1} X_{1D}  + h_{21}\sqrt{\delta_1} X_{1G}+h_{22}\sqrt{1-\delta_2} X_{2Gc} + h_{22}\sqrt{\delta_2} X_{2Gp}+Z_2.
\end{align*}
The achievable region in~\eqref{eq:inner bound} with  $Q=\emptyset, U_2 = X_{2Gc}$ reduces to
\begin{align*}
   R_1 & \leq I(X_1;Y_1|X_{2Gc}) \notag
\\&=h(Y_1|X_{2Gc}) -h(Y_1|X_1,X_{2Gc})
\\&=h(h_{11}\sqrt{1-\delta_1} X_{1D}  + h_{11}\sqrt{\delta_1} X_{1G}+ h_{12}\sqrt{\delta_2} X_{2Gp}+Z_1) 
\\&-h(h_{12}\sqrt{\delta_2} X_{2Gp}+Z_1)
\\&=h\left(\sqrt{\frac{|h_{11}|^2(1-\delta_1)}{1+|h_{11}|^2\delta_1 + |h_{12}|^2\delta_2}} X_{1D} +Z_1\right) -h(Z_1)
\\&+\mug\left(|h_{11}|^2\delta_1 + |h_{12}|^2\delta_2\right)
   -\mug\left(|h_{12}|^2\delta_2\right); 
\end{align*}
therefore, by Theorem~\ref{thm:lowbound}, we can further lower bound the rate of user~1 as
\begin{align*}
R_1 \leq \mud\left(N, \frac{|h_{11}|^2(1-\delta_1)}{1+|h_{11}|^2\delta_1 + |h_{12}|^2\delta_2}  \right)
        +\mug\left(\frac{|h_{11}|^2\delta_1}{1 + |h_{12}|^2\delta_2}\right),
\end{align*}
thus proving~\eqref{eq1:ach2:withScheme2}.

For the rate of user~2 we have
\begin{align*}
 R_2&\leq I(X_2;Y_2)
\\&=h\left(h_{21}\sqrt{1-\delta_1} X_{1D}  + h_{21}\sqrt{\delta_1} X_{1G}+h_{22}\sqrt{1-\delta_2} X_{2Gc} + h_{22}\sqrt{\delta_2} X_{2Gp}+Z_2\right)
\\&-h\left(h_{21}\sqrt{1-\delta_1} X_{1D}  + h_{21}\sqrt{\delta_1} X_{1G}+Z_2\right)
\\&=h\left(\sqrt{\frac{|h_{21}|^2(1-\delta_1)}{1+|h_{21}|^2\delta_1+|h_{22}|^2}} X_{1D}  +Z_2\right)-h(Z_2) +\mug\left(|h_{21}|^2\delta_1+|h_{22}|^2\right)
\\&-h\left(\sqrt{\frac{|h_{21}|^2(1-\delta_1)}{1+|h_{21}|^2\delta_1}} X_{1D}  +Z_2\right)+h(Z_2)-\mug\left(|h_{21}|^2\delta_1\right)
\end{align*}
therefore, by Theorem~\ref{thm:lowbound}, we can further lower bound the rate of user~2 as
\begin{align*}
R_2
 &\leq \mud\left(N, \frac{|h_{21}|^2(1-\delta_1)}{1+|h_{21}|^2\delta_1+|h_{22}|^2} \right)
   +\mug\left(\frac{|h_{22}|^2}{1+|h_{21}|^2\delta_1}\right)
\\&-\mug  \left( \min\left(N^2-1,\frac{|h_{21}|^2(1-\delta_1)}{1+|h_{21}|^2\delta_1}\right)  \right)
\end{align*}
thus proving~\eqref{eq2:ach2:withScheme2}.

Finally for the sum-rate we have
\begin{align*}
 R_1+R_2 &\leq I(X_1,X_{2Gc};Y_1)+I(X_2;Y_2|X_{2Gc})
\\&=h(h_{11}\sqrt{1-\delta_1} X_{1D}  + h_{11}\sqrt{\delta_1} X_{1G}+ h_{12}\sqrt{1-\delta_2} X_{2Gc}+ h_{12}\sqrt{\delta_2} X_{2Gp}+Z_1)
\\&-h(h_{12}\sqrt{\delta_2} X_{2Gp}+Z_1)
\\&+h(h_{21}\sqrt{1-\delta_1} X_{1D}  + h_{21}\sqrt{\delta_1} X_{1G} + h_{22}\sqrt{\delta_2} X_{2Gp}+Z_2)
\\&-h(h_{21}\sqrt{1-\delta_1} X_{1D}  + h_{21}\sqrt{\delta_1} X_{1G} +Z_2)'
\end{align*}
therefore, by Theorem~\ref{thm:lowbound}, we can further lower bound the sum-rate as
\begin{align*}
 R_1+R_2 &\leq\mud\left(N, \frac{|h_{11}|^2(1-\delta_1)}{1+|h_{11}|^2\delta_1 + |h_{12}|^2}  \right)+\mug\left(|h_{11}|^2\delta_1 + |h_{12}|^2 \right)
\\&-\mug\left(|h_{12}|^2\delta_2 \right)
\\&+\mud\left(N, \frac{|h_{21}|^2(1-\delta_1)}{1+|h_{21}|^2\delta_1 + |h_{22}|^2\delta_2}  \right)+\mug\left(|h_{21}|^2\delta_1 + |h_{22}|^2\delta_2 \right)
\\&- \mug\left( \min\left(N^2-1, \frac{|h_{21}|^2(1-\delta_1)}{1+|h_{21}|^2\delta_1}\right)  \right)-\mug\left(|h_{21}|^2\delta_1 \right)
\end{align*}
thus proving~\eqref{eq3:ach2:withScheme2}.

\begin{rem}\label{rem:scheme2+etw power split}
For future use, we specialized the derived achievable rate region for the power splits $\delta_1=\frac{1}{1+|h_{21}|^2}$ and $\delta_2=\frac{1}{1+|h_{12}|^2}$ inspired by \cite{etkin_tse_wang}; we thus have that the following region is achievable for any $N\in\mathbb{N}$
\begin{subequations}
\begin{align}
R_1 &\le   
    \mud\left(N, \frac{|h_{11}|^2{{a}} }{1+\frac{|h_{11}|^2}{1+|h_{21}|^2}  + {{b}} }  \right)
   +\mug\left(\frac{\frac{|h_{11}|^2}{1+|h_{21}|^2}}{1+{{b}}}\right),
\label{eq1:ach2:etw}\\
R_2 & \le
    \mud\left(N, \frac{|h_{21}|^2{{a}} }{1+{{a}} +|h_{22}|^2} \right)
   +\mug\left(\frac{|h_{22}|^2}{1+{{a}} }\right) \notag
\\&-\mug\left(\min\left(N^2-1,|h_{21}|^2 {{a}}\right)  \right),
\label{eq2:ach2:etw}\\
R_1+R_2 & \le
    \mud\left(N, \frac{|h_{11}|^2{{a}} }{1+\frac{|h_{11}|^2}{1+|h_{21}|^2}  + |h_{12}|^2}  \right)
   +\mug\left(\frac{|h_{11}|^2}{1+|h_{21}|^2}  + |h_{12}|^2 \right)
   -\mug\left({{b}}  \right) \notag
\\&+\mud\left(N, \frac{|h_{21}|^2{{a}} }{1+{{a}}  + \frac{|h_{22}|^2}{1+|h_{12}|^2} }  \right)\notag
   +\mug\left(\frac{\frac{|h_{22}|^2}{1+|h_{12}|^2} }{1+{{a}}}\right) \notag
\\&-\mug\left( \min\left(N^2-1, |h_{21}|^2 {{a}} \right)  \right).
\label{eq3:ach2:etw}
\end{align}
\label{eqall:ach2:etw}
\end{subequations}
where $a := \frac{|h_{21}|^2}{1+|h_{21}|^2} \in[0,1]$ and $b :=  \frac{|h_{12}|^2}{1+|h_{12}|^2} \in[0,1]$.

In the symmetric case the region in~\eqref{eqall:ach2:etw} is further lower bounded by
\begin{subequations}
\begin{align}
R_1 &\le   
    \mug\left(\min\left(N^2-1,\frac{\snr \cdot \inr }{1+\snr+2\inr}  \right)\right) - \frac{1}{2}\log\left(\frac{\pi\eu}{3}\right) 
   +\mug\left(\frac{\snr}{1+2\inr}\right),
\label{eq1:ach2:etw sym}\\
R_2 & \le
   \mug\left(\min\left(N^2-1,\frac{\inr^2}{(1+\inr)(1+\snr)+\inr} \right)\right) - \frac{1}{2}\log\left(\frac{\pi\eu}{3}\right) 
   \notag
\\&+\mug\left(\snr\frac{1}{2}\right)
   -\mug\left(\min\left(N^2-1,\frac{\inr^2}{1+2\inr}\right)  \right),
\label{eq2:ach2:etw sym}\\
R_1+R_2 & \le
    \mug\left(\min\left(N^2-1,\frac{\snr \cdot \inr }{(1+\inr)^2+\snr}\right)\right) - \frac{1}{2}\log\left(\frac{\pi\eu}{3}\right) \notag
\\&+\mug\left(\inr+\frac{\snr}{1+\inr}\right)
   -\mug\left(\frac{\inr}{1+\inr}\right) \notag
\\&+\mug\left(\min\left(N^2-1,\frac{\inr^2}{1+\snr+2\inr} \right)\right) - \frac{1}{2}\log\left(\frac{\pi\eu}{3}\right)      \notag
\\&+\mug\left(\frac{\snr}{1+2\inr}\right)-\mug\left( \min\left(N^2-1, \frac{\inr^2}{1+2\inr}\right)  \right).
\label{eq3:ach2:etw sym}
\end{align}
\label{eqall:ach2:etw sym}
\end{subequations}
\end{rem}

\section{Gap derivation for the moderately weak interference regime}
\label{app:gap moderately week tedious}
In order to show achievability to within a constant gap of the outer bound in~\eqref{eq:etw outer bound:all} by means of the achievable region in~\eqref{eqall:ach2:etw sym weak} (a further lower bound to the region in~\eqref{eqall:ach2:etw sym}), we distinguish two cases.

\subsection*{CASE 1 (regime corresponding to $\alpha\in[2/3,1]$ in Fig.~\ref{fig:gDoFachModerate})}
Assume that the sum-rate in eq.\eqref{eq3:ach2:etw sym weak} is redundant;
under this condition we match the corner point of the rectangular achievable region,
given by $(R_1,R_2)=(\text{eq.\eqref{eq1:ach2:etw sym weak}, \ eq.\eqref{eq2:ach2:etw sym weak}})$, to
\begin{subequations}
\begin{align}
\mathcal{R}_{\rm out}^{\text{(G-IC mod P1)}} : \quad
R_1 &= \mug\left(\inr+\frac{\snr}{1+\inr} \right),
\label{eq:CornerPoint1Moderate r1}
\\
R_2 &= \mug\left(\snr\right)-\mug\left(\inr\right) +\mug(\inr+\snr)-\mug\left(\inr+\frac{\snr}{1+\inr} \right),
\label{eq:CornerPoint1Moderate r2}
\end{align}
\label{eq:CornerPoint1Moderate}
\end{subequations}
and
\begin{subequations}
\begin{align}
\mathcal{R}_{\rm out}^{\text{(G-IC mod P2)}} : \quad
R_1 &= \mug\left(\snr\right)-\mug\left(\inr\right) +\mug(\inr+\snr)-\mug\left(\inr+\frac{\snr}{1+\inr} \right),
\label{eq:CornerPoint2Moderate r1}
\\
R_2 &= \mug\left(\inr+\frac{\snr}{1+\inr} \right),
\label{eq:CornerPoint2Moderate r2}
\end{align}
\label{eq:CornerPoint2Moderate}
\end{subequations}
which were obtained from the intersection of the sum-rate outer bound in~\eqref{eq:etw outer bound:r1+r2 kra} with either~\eqref{eq:etw outer bound:2r1+r2} or~\eqref{eq:etw outer bound:r1+2r2}. 
In particular,
for the corner point in~\eqref{eq:CornerPoint1Moderate} we use $x$ in~\eqref{eq:x for NpointCornerPoint1weak} (which corresponds to $N$ in~\eqref{eq:NpointCornerPoint1weak}), and 
for the corner point in~\eqref{eq:CornerPoint2Moderate} we use $x$ in~\eqref{eq:x for NpointCornerPoint2weak} (which corresponds to $N$ in~\eqref{eq:NpointCornerPoint2weak}).

The gap is readily computed as follows:
for the corner point in~\eqref{eq:CornerPoint1Moderate} we have
\begin{align*}
\Delta_1 &= \text{eq.\eqref{eq:CornerPoint1Moderate r1}-eq.\eqref{eq1:ach2:etw sym weak}}|_{\text{$x$ in~\eqref{eq:x for NpointCornerPoint1weak}}}
\\&\leq 
     \mug\left(\inr+\frac{\snr}{1+\inr} \right)
    -\mug\left(\frac{\snr}{1+2\inr}\right)
    -\mug\left(\frac{\inr^2}{1+\snr+2\inr}\right)
\\&+ 
     \frac{1}{2}\log(4)
    +
    \frac{1}{2}\log\left(\frac{\pi\eu}{3}\right)
\\&\leq
     \frac{1}{2}\log\left(2\right)
    +\frac{1}{2}\log\left(4\right)
    +\frac{1}{2}\log\left(\frac{\pi\eu}{3}\right)
    =\frac{1}{2}\log\left(\frac{8\pi\eu}{3}\right),
\end{align*}
and
\begin{align*}
\Delta_2 &= \text{eq.\eqref{eq:CornerPoint1Moderate r2}-eq.\eqref{eq2:ach2:etw sym weak}}|_{\text{$x$ in~\eqref{eq:x for NpointCornerPoint1weak}}}
\\&\leq
     \mug\left(\snr\right)
    -\mug\left(\inr\right)
    +\mug(\inr+\snr)
    -\mug\left(\inr+\frac{\snr}{1+\inr} \right)
\\& -\mug\left(\frac{\snr}{2}\right)
    +\mug\left(\frac{\inr^2}{1+\snr+2\inr}\right)
    { -\mug\left(\frac{\inr^2}{(1+\inr)(1+\snr)+\inr}\right)} 
    { +\frac{1}{2}\log\left(\frac{\pi\eu}{3}\right)}
\\&\leq
      { \frac{1}{2}\log\left(2\right)}
     +{ \frac{1}{2}\log\left(\frac{\pi\eu}{3}\right)}
     =\frac{1}{2}\log\left(\frac{2\pi\eu}{3}\right), \ \text{since $\inr \leq \snr$ in weak interfernce};
\end{align*}
while for the corner point in~\eqref{eq:CornerPoint2Moderate} we have
\begin{align*}
\Delta_1 &= \text{eq.\eqref{eq:CornerPoint2Moderate r1}-eq.\eqref{eq1:ach2:etw sym weak}}|_{\text{$x$ in~\eqref{eq:x for NpointCornerPoint2weak}}}
\\&\leq
     \mug\left(\snr\right)
    -\mug\left(\inr\right)
    +\mug(\inr+\snr)
    -\mug\left(\inr+\frac{\snr}{1+\inr} \right)
\\& -\mug\left(\frac{\snr}{1+2\inr}\right)
    -\mug\left(\frac{\snr \cdot \inr }{(1+\inr)^2+\snr}\right)
    +\frac{1}{2}\log\left(4\right)
    +\frac{1}{2}\log\left(\frac{\pi\eu}{3}\right)
\\&\leq
     \frac{1}{2}\log\left(2\right)
    +\frac{1}{2}\log\left(4\right)
    +\frac{1}{2}\log\left(\frac{\pi\eu}{3}\right)
    =\frac{1}{2}\log\left(\frac{8\pi\eu}{3}\right),
\end{align*}
and
\begin{align*}
\Delta_2 &= \text{eq.\eqref{eq:CornerPoint2Moderate r2}-eq.\eqref{eq2:ach2:etw sym weak}}|_{\text{$x$ in~\eqref{eq:x for NpointCornerPoint2weak}}}
\\&\leq
     \mug\left(\inr+\frac{\snr}{1+\inr} \right)
    -\mug\left(\frac{\snr}{2}\right)
    +\mug\left(\frac{\snr \cdot \inr }{(1+\inr)^2+\snr}\right)
\\&
     { -\mug\left(\frac{\inr^2}{(1+\inr)(1+\snr)+\inr}\right)}
    +{ \frac{1}{2}\log\left(\frac{\pi\eu}{3}\right)}
\\&\leq
     { \frac{1}{2}\log\left(2\right)}
    +{ \frac{1}{2}\log\left(\frac{\pi\eu}{3}\right)}
     =\frac{1}{2}\log\left(\frac{2\pi\eu}{3}\right), \ \text{since $\inr \leq \snr$ in weak interference}.
\end{align*}

\subsection*{CASE 2 (regime corresponding to $\alpha\in[1/2,2/3]$ in Fig.~\ref{fig:gDoFachWeak})}
Assume that the sum-rate in~\eqref{eqall:ach2:etw sym weak} is not redundant, that is after simple algebraic manipulation,  
\begin{align*}
&1+\min\left(x|_{\text{$x$ in~\eqref{eq:x for NpointCornerPoint1weak}}},
             x|_{\text{$x$ in~\eqref{eq:x for NpointCornerPoint2weak}}}\right)
\notag\\&<  \underbrace{
    \frac{(1+2\inr)(1+\frac{\snr}{2})}{ (1+\inr)(1+\snr)+\inr}}_{\in[ 0.7358 ,1] \ \text{for} \ \inr \leq \snr \leq \inr(1+\inr) {  \text{ see Appendix~\ref{sec:min:function}}}} 
    \cdot \quad
    \underbrace{\frac{(1+\inr)(1+\inr+\snr)}{(1+\inr)^2+\snr}}_{=1+x|_{\text{$x$ in~\eqref{eq:x for NpointCornerPoint2weak}}}},
\end{align*}
which implies
\begin{align}
x|_{\text{$x$ in~\eqref{eq:x for NpointCornerPoint1weak}}} \leq
x|_{\text{$x$ in~\eqref{eq:x for NpointCornerPoint2weak}}}.
\label{eq:condition redundancy sumrate scheme2}
\end{align}

Under the condition in~\eqref{eq:condition redundancy sumrate scheme2} we match one of the corner point of the pentagon-shaped achievable region in~\eqref{eqall:ach2:etw sym weak} to 
\begin{subequations}
\begin{align}
\mathcal{R}_{\rm out}^{\text{(G-IC weak P1)}} :
R_1 &= 3\mug\left(\inr+\frac{\snr}{1+\inr}\right) -\mug\left(\snr+\inr\right)-\mug\left(\snr\right)+\mug\left(\inr\right), 
\label{eq:CornerPoint1Weak r1}\\
R_2 &= \mug\left(\snr\right)-\mug\left(\inr\right)+\mug\left(\snr+\inr\right)-\mug\left(\inr+\frac{\snr}{1+\inr} \right),
\label{eq:CornerPoint1Weak r2}
\end{align}
\label{eq:CornerPoint1Weak}
\end{subequations}
and 
\begin{subequations}
\begin{align}
\mathcal{R}_{\rm out}^{\text{(G-IC weak P2)}} :
R_1 &= \mug\left(\snr\right)-\mug\left(\inr\right)+\mug\left(\snr+\inr\right)-\mug\left(\inr+\frac{\snr}{1+\inr} \right),
\label{eq:CornerPoint2Weak r1}\\
R_2 &= 3\mug\left(\inr+\frac{\snr}{1+\inr}\right) -\mug\left(\snr+\inr\right)-\mug\left(\snr\right)+\mug\left(\inr\right), 
\label{eq:CornerPoint2Weak r2}
\end{align}
\label{eq:CornerPoint2Weak}
\end{subequations}
which were obtained from the intersection of the sum-rate outer bound in~\eqref{eq:etw outer bound:r1+r2 etw} with either~\eqref{eq:etw outer bound:2r1+r2} or~\eqref{eq:etw outer bound:r1+2r2}. 
In particular,
for the corner point in~\eqref{eq:CornerPoint1Weak} we use $x$ in~\eqref{eq:x for NpointCornerPoint1weak} (which corresponds to $N$ in~\eqref{eq:NpointCornerPoint1weak}), and 
for the corner point in~\eqref{eq:CornerPoint2Weak} we use $x$ in~\eqref{eq:x for NpointCornerPoint2weak} (which corresponds to $N$ in~\eqref{eq:NpointCornerPoint2weak}).

The gap is readily computed as follows:
for the corner point in~\eqref{eq:CornerPoint1Weak} we have
\begin{align*}
\Delta_1 &= \text{eq.\eqref{eq:CornerPoint1Weak r1}-\Big(eq.\eqref{eq3:ach2:etw sym weak}-eq.\eqref{eq2:ach2:etw sym weak}\Big)}|_{\text{$x$ in~\eqref{eq:x for NpointCornerPoint1weak}}}
\\&\leq 
   2\mug\left(\inr+\frac{\snr}{1+\inr}\right)
   -\mug\left(\snr+\inr\right)
   -\mug\left(\snr\right)
   +\mug\left(\inr\right)
   +\mug\left(\frac{\inr}{1+\inr}\right)
\\&
   -\mug\left(\frac{\snr}{1+2\inr}\right)
   +\mug\left(\frac{\snr}{2}\right)
+\mug\left(\frac{\inr^2}{(1+\inr)(1+\snr)+\inr}\right)
  \\& -2\mug\left(\frac{\inr^2}{1+\snr+2\inr}\right)
   + \frac{1}{2}\log\left(4\right)
   +\frac{1}{2}\log\left(\frac{\pi\eu}{3}\right) 
\\& = \frac{1}{2} \log\left ( \frac{\left(\frac{\snr}{2} + 1\right)\, \left(\frac{\inr}{\inr + 1} + 1\right)\, \left(\frac{{\inr}^2}{\inr + \left(\inr + 1\right)\, \left(\snr + 1\right)} + 1\right)\, \left(\inr + 1\right)\, {\left(\inr + \frac{\snr}{\inr + 1} + 1\right)}^2}{{\left(\frac{{\inr}^2}{2\, \inr + \snr + 1} + 1\right)}^2\, \left(\frac{\snr}{2\, \inr + 1} + 1\right)\, \left(\snr + 1\right)\, \left(\inr + \snr + 1\right)} \right)
\\& +\frac{1}{2}\log\left(4\right)
   +\frac{1}{2}\log\left(\frac{\pi\eu}{3}\right) \\
   &= \frac{1}{2} \log \left(\frac{{\left(2\, \inr + 1\right)}^2\, \left(\frac{\snr}{2} + 1\right)\, \left(2\, \inr + \snr + 1\right)}{\left(\inr + 1\right)\, \left(\snr + 1\right)\, \left(2\, \inr + \snr + \inr\, \snr + 1\right)} \right)+\frac{1}{2}\log\left(\frac{\pi\eu}{3}\right) \\
   &\leq \frac{1}{2}\log(6)+\frac{1}{2}\log\left(4\right) +\frac{1}{2}\log\left(\frac{\pi\eu}{3}\right) =\frac{1}{2}\log\left(8\pi\eu\right)
\end{align*}
and
\begin{align*}
\Delta_2 &= \text{eq.\eqref{eq:CornerPoint1Weak r2}-eq.\eqref{eq2:ach2:etw sym weak}}|_{\text{$x$ in~\eqref{eq:x for NpointCornerPoint1weak}}}
\\&\leq
  \mug\left(\snr\right)
 -\mug\left(\inr\right)
 +\mug\left(\snr+\inr\right)
 -\mug\left(\inr+\frac{\snr}{1+\inr} \right)
 -\mug\left(\frac{\snr}{2}\right)
\\& +\mug\left(\frac{\inr^2}{1+\snr+2\inr}\right)
-\mug\left(\frac{\inr^2}{(1+\inr)(1+\snr)+\inr}\right)
 +  \frac{1}{2}\log\left(\frac{\pi\eu}{3}\right)
\\&=
  \frac{1}{2}\log\left(\frac{\left(\frac{{\inr}^2}{2\, \inr + \snr + 1} + 1\right)\, \left(\snr + 1\right)\, \left(\inr + \snr + 1\right)}{\left(\frac{\snr}{2} + 1\right)\, \left(\frac{{\inr}^2}{\inr + \left(\inr + 1\right)\, \left(\snr + 1\right)} + 1\right)\, \left(\inr + 1\right)\, \left(\inr + \frac{\snr}{\inr + 1} + 1\right)}\right)
  \\&+ \frac{1}{2}\log\left(\frac{\pi\eu}{3}\right)
  \\&=\frac{1}{2} \log \left( \frac{2\, \left(\snr + 1\right)\, \left(2\, \inr + \snr + \inr\, \snr + 1\right)}{\left(\inr + 1\right)\, \left(\snr + 2\right)\, \left(2\, \inr + \snr + 1\right)} \right)+\frac{1}{2}\log\left(\frac{\pi\eu}{3}\right)
  \\& \le \frac{1}{2}\log\left(2\right)+\frac{1}{2}\log\left(\frac{\pi\eu}{3}\right)=\frac{1}{2}\log\left(\frac{2\pi\eu}{3}\right) ,
\end{align*}
while for the corner point in~\eqref{eq:CornerPoint2Weak} we have
\begin{align*}
\Delta_1 &= \text{eq.\eqref{eq:CornerPoint2Weak r1}-eq.\eqref{eq1:ach2:etw sym weak}}|_{\text{$x$ in~\eqref{eq:x for NpointCornerPoint2weak}}}
\\&\leq
   \mug\left(\snr\right)
  -\mug\left(\inr\right)
  +\mug\left(\snr+\inr\right)
  -\mug\left(\inr+\frac{\snr}{1+\inr} \right)
  -\mug\left(\frac{\snr}{1+2\inr}\right)
 \\& -\mug\left(\frac{\snr \cdot \inr }{(1+\inr)^2+\snr}\right)
   +\frac{1}{2}\log\left(\frac{\pi\eu}{3}\right)+\frac{1}{2}\log(4)
\\&=
   \frac{1}{2}\log\left(\frac{\left(\snr + 1\right)\, \left(\inr + \snr + 1\right)}{\left(\frac{\snr}{2\, \inr + 1} + 1\right)\, \left(\inr + 1\right)\, \left(\frac{\inr\, \snr}{\snr + {\left(\inr + 1\right)}^2} + 1\right)\, \left(\inr + \frac{\snr}{\inr + 1} + 1\right)}\right)
 \\& +\frac{1}{2}\log\left(4\right)
  +\frac{1}{2}\log\left(\frac{\pi\eu}{3}\right)
  \\ &=\frac{1}{2}\log\left(\frac{\left(2\, \inr + 1\right)\, \left(\snr + 1\right)}{\left(\inr + 1\right)\, \left(2\, \inr + \snr + 1\right)}\right) +\frac{1}{2}\log\left(4\right)
  +\frac{1}{2}\log\left(\frac{\pi\eu}{3}\right)\\
 & \le \frac{1}{2}\log\left(2\right)+\frac{1}{2}\log\left(4\right)+\frac{1}{2}\log\left(\frac{\pi\eu}{3}\right) =\frac{1}{2}\log\left(\frac{4\pi\eu}{3}\right),
\end{align*}
and
\begin{align*}
\Delta_2 &= \text{eq.\eqref{eq:CornerPoint2Weak r2}-
  \Big(eq.\eqref{eq3:ach2:etw sym weak}-eq.\eqref{eq1:ach2:etw sym weak}\Big)}|_{\text{$x$ in~\eqref{eq:x for NpointCornerPoint2weak}}}
\\&\leq
   2\mug\left(\inr+\frac{\snr}{1+\inr}\right)
   -\mug\left(\snr+\inr\right)
   -\mug\left(\snr\right)
   +\mug\left(\inr\right)
   +\mug\left(\frac{\inr}{1+\inr}\right)
   \\&+\frac{1}{2}\log\left(\frac{\pi\eu}{3}\right)
\\
& = \frac{1}{2}\log \left( \frac{(1+2\inr)((1+\inr)^2+\snr)}{(1+\inr)^2(1+\snr)(1+\inr+\snr)} \right)  +\frac{1}{2}\log\left(\frac{\pi\eu}{3}\right)
\\& \le 0+\frac{1}{2}\log\left(\frac{\pi\eu}{3}\right)=\frac{1}{2}\log\left(\frac{\pi\eu}{3}\right)
\end{align*}
This concludes the proof.

{ 
\section{Minimum of a function}
\label{sec:min:function}
The minimum of the function
\[
f(x,y)=\frac{(1+2y)(1+\frac{x}{2})}{(1+y)(1+x)+y}, \quad \text{for $(x,y)\in\mathbb{R}^2_+$ such that} \quad 1 \le y \le x \le y(1+y), 
\]
is found by first taking the partial derivative  with respect to $x$, given my $\frac{\partial f}{\partial x}=-\frac{2y^2+7y+3}{2(2x+y+xy+1)^2}$ which is easily seen to be monotone decreasing in $x$ therefore attaining the minimum
\[
f(y(1+y),y)=\frac{2 y^3 + 3 y^2 + 5 y + 2}{2y^3 + 6 y^2 + 6y + 2}, \quad \text{for} \quad 1 \le y.
\]
Now by taking the partial derivative with respect to $y$, given by $\frac{\partial f}{\partial y}= \frac{\left(  3 y^2 - 4 y -1\right)}{2 {\left(y + 1\right)}^4}$ and setting it equal to zero 
we see that the minimum occurs at $y=\frac{\sqrt{7}+2}{3}$. Hence, the minimum of the function occurs at $f\left(\frac{\sqrt{7}+2}{3} \left(1+\frac{\sqrt{7}+2}{3} \right),\frac{\sqrt{7}+2}{3}\right)=0.7359$.
Conditions on the second derivatives can be easily checked to verify that indeed the claim stationary point is a global minimum (even easier still, by plotting the function for example with Matlab). 
}

\bibliography{refs}
\bibliographystyle{IEEEtran}

\end{document}